\documentclass[reprint,twocolumn,showpacs,superscriptaddress]{revtex4-1}

\usepackage{epsfig}
\usepackage{amsmath,amssymb}
\usepackage{graphicx}
\usepackage{hyperref}
\hypersetup{colorlinks=true,urlcolor=blue,citecolor=blue,linkcolor=blue,breaklinks=true}
\usepackage[noabbrev]{cleveref}
\usepackage{adjustbox}
\usepackage{floatrow}
\usepackage{footnote}
\usepackage{grffile}
\usepackage{color}
\usepackage{xcolor}
\usepackage{dsfont}
\usepackage{mathtools}
\usepackage{verbatim}
\usepackage{bbold}
\usepackage{hhline}
\usepackage{textcomp}

\allowdisplaybreaks

\newcommand{\cw}{\columnwidth}

\begin{document}
\title{Quantitative functional renormalization-group description of the two-dimensional Hubbard model}

\newcommand{\TUVienna}{\affiliation{Institute of Solid State Physics, Vienna University of Technology, 1040 Vienna, Austria}}
\newcommand{\UniTueb}{\affiliation{Institut f\"ur Theoretische Physik and Center for Quantum Science, Universit\"at T\"ubingen, Auf der Morgenstelle 14, 72076 T\"ubingen, Germany}}
\newcommand{\RWTH}{\affiliation{Institute for Theoretical Solid State Physics, RWTH Aachen University, 52056 Aachen, Germany}}
\newcommand{\LMUold}{\affiliation{Physics Department, Arnold Sommerfeld Center for Theoretical Physics, and Center for NanoScience, Ludwig-Maximilians-Universität München}}
\newcommand{\LMU}{\affiliation{Arnold Sommerfeld Center for Theoretical Physics, 
Center for NanoScience,\looseness=-1\, and Munich Center for \\ Quantum Science and Technology,\looseness=-2\, Ludwig-Maximilians-Universit\"at M\"unchen, 80333 Munich, Germany}}
\newcommand{\Simons}{\affiliation{Center for Computational Quantum Physics, Flatiron Institute, New York, New York 10010, USA}}
\newcommand{\WilliamMary}{\affiliation{Department of Physics, College of William and Mary, Williamsburg, Virginia 23187, USA}}
 \newcommand{\Jara}{\affiliation{JARA-FIT, JARA-HPC, J\"ulich Aachen Research Alliance, 52425 J\"ulich, Germany}}
 
\author{Cornelia Hille}       \UniTueb

\author{Fabian B.~Kugler}         \LMU

\author{Christian J.~Eckhardt} 		\TUVienna\RWTH\Jara

\author{Yuan-Yao He}    \Simons\WilliamMary

\author{Anna Kauch} 		\TUVienna

\author{Carsten Honerkamp} 	\RWTH\Jara

\author{Alessandro Toschi}  	\TUVienna

\author{Sabine Andergassen} 	\UniTueb

\begin{abstract}
Using a forefront algorithmic implementation of the functional renormalization group (fRG) for interacting fermions on two-dimensional lattices, we provide a detailed analysis of its quantitative reliability for the Hubbard model. In particular, we show that the recently introduced multiloop extension of the fRG flow equations for the self-energy and two-particle vertex allows for a precise match with the parquet approximation {\sl also} for two-dimensional lattice problems. The refinement with respect to previous fRG-based computation schemes relies on an accurate treatment of the frequency and momentum dependences of the two-particle vertex, which combines a proper inclusion of the high-frequency asymptotics with the so-called `truncated unity' fRG for the momentum dependence. The adoption of the latter scheme requires, as an essential step, a consistent modification of the flow equation of the self-energy. We quantitatively compare our fRG results for the self-energy and momentum-dependent susceptibilities and the corresponding solution of the parquet approximation to determinant quantum Monte Carlo data, demonstrating that the fRG is remarkably accurate up to moderate interaction strengths. The presented methodological improvements illustrate how fRG flows can be brought to a quantitative level for two-dimensional problems, providing a solid basis for the application to more general systems.
\end{abstract}

\maketitle

\section{Introduction}
\label{sec:introduction}

Renormalization group (RG) methods have a long history in theoretical physics, ranging from a way to treat divergences in quantum field theories \cite{Weinberg1995}, critical phenomena \cite{Wilson,Wegner}, and quantum impurity problems \cite{Anderson1970,Wilson1975} to current attempts to elucidate deep learning algorithms by physics \cite{SchwabMehta}. 
In general, RG methods connect specific quantities of a theory, such as coupling constants or correlation functions, at a given scale with those at another scale via differential equations. This leads to a flow of these quantities which under appropriate circumstances distills out the dominating, and to some degree universal, properties of the system.   

The origins of the RG for electron lattice models date back to the second decade of the cuprate high-temperature superconductors more than 20 years ago \cite{Zanchi1996,Zanchi1998,Halboth,Tsai,hsfr}. Here, the RG was utilized as a tool to deal with competing ordering tendencies in the Hubbard model (which are also seen in the cuprates) and as a method to understand in principle the stability of the Landau--Fermi-liquid state \cite{Salmhofer1998}. While the aptness of RG schemes in competing-order situations had already been known from impurity \cite{Anderson1970} and one-dimensional models \cite{Solyom1979}, the systematic and versatile functional RG schemes  \cite{Polchinski1984,Wetterich1993,Salmhofer1998,Kopietz2010} turned out to be advantageous in the study of two-dimensional (2D) lattice models such as the Hubbard model, beyond more general considerations \cite{Polchinski1992,Shankar1994}. 

The name functional RG (fRG) can be understood as having a twofold reason: On the one hand, the RG flow is derived from an exact flow equation for a generating functional of the theory when a (suitably chosen) parameter $\Lambda $ in the free, quadratic part of the action is changed. On the other hand, one usually investigates the flow of continuous functions of variables $k$ such as wavevectors and frequencies with $\Lambda$, i.e., one deals with differential equations in $\Lambda$ for functions $f_\Lambda(k)=f(\Lambda,k)$ of $k$ and $\Lambda$. This marks a difference with respect to the conventional RG, where only a small finite number of constants is flowing. The main objects of interest in the fRG flows we discuss here are the electron single-particle self-energy $\Sigma_{\Lambda} (k_1,k_2)$ and the two-particle (interaction) vertex $V_\Lambda(k_1,k_2,k_3,k_4)$. While the self-energy determines the changes of the single-particle excitations due to the mutual interactions, the two-particle correlation functions and collective properties are mostly controlled by the two-particle scattering processes, the so-called `vertex correction' terms. A dominant role played by the latter clearly emerged from model \cite{Vilk1997,Bergeron2011,Rohringer2012,Hafermann2014,Janis2014,Vucicevic2019,Reza2019,Kauch2020,Springer2020} and realistic \cite{Toschi2012,Liu2012,Galler2015,Hausoel2017,Watzenboeck2020} calculations of correlated materials.

Numerically, the costly parts are the evaluation of the fRG differential flow equations, usually determined by Feynman-type loop
diagrams, as well as the high number of components of the flowing functions such as $V_\Lambda(k_1,k_2,k_3,k_4)$.  

Initially, the fRG in the form used here was developed and employed in 2D Hubbard models in view of high-$T_c$ cuprates. First works using so-called `$N$-patch techniques' to resolve the emerging wavevector dependence of the flowing interaction focused on the exploration of the leading ordering tendencies of the model \cite{Zanchi1996,Zanchi1998,Halboth,Tsai} and understanding similarities to one-dimensional models. This included the question whether the pseudogap of the cuprates is foreshadowed in the fRG flow of weak to moderately coupled Hubbard models \cite{hsfr}. Parallel developments were conducted in the field of inhomogeneous one-dimensional systems \cite{Andergassen2004,Andergassen2006b}. Already at early stages, they were able to perform some quantitative benchmarking with exact numerical techniques (density matrix renormalization group), and to include partial self-energy corrections in the fRG. For 2D models, incorporating the fRG flow of the self-energy was technically more difficult for a number of reasons \cite{hsfr}, but works indicating the opening of pseudogaps in the 2D flows came out a few years later \cite{KataninKampf,Rohe2005}. However, at that time, few attempts were made to assess the quantitative precision of the method in comparison with other theoretical methods. Rather, the $N$-patch fRG techniques were used in numerous applications to broad classes of experimentally relevant material systems, such as the iron superconductors \cite{FaWang2009,Platt2013}, graphene \cite{Honerkamp2008,Kiesel2012,Wang2012}, and in the search for interaction-induced topological states  \cite{Raghu2008,Scherer2Honerkamp}. There, the fRG was mainly used as a flexible tool to explore ground-state phase diagrams over a wide range of parameters. 

Besides these applications, the formal development saw a steady evolution. 
Recently, the long-standing challenge of relating fRG schemes to the parquet approach was addressed \cite{Kugler2018a,Kugler2018b}. Within the so-called `multiloop extension' of the fRG, all higher-loop contributions to the flow of parquet diagrams are accounted for with their exact weight.
Conveniently, the effort only grows linearly with the loop order.
The equivalence between the parquet approximation (PA) and multiloop fRG has been rigorously established at the level of spinless \cite{Kugler2018a} and spinful \cite{Chalupa2020} impurity models. 
Its numerical verification in the 2D Hubbard model poses additional challenges, due to the necessity of additionally treating the 2D momentum variables.
Specifically, we will demonstrate how it is possible to render the numerical effort for a quantitative treatment of the fRG flow manageable in 2D, by exploiting the multichannel decomposition \cite{Karrasch2008} of the interaction vertex in combination with form-factor expansions \cite{Husemann2009,Wang2012}. This frees resources to incorporate frequency dependencies and self-energy corrections in various cases  \cite{Giering,Eberlein2015}, and to conduct flows into symmetry-broken states  \cite{Eberlein2014,Maier2014}.
Strongly inspired by earlier channel-decomposition schemes \cite{Husemann2009} and the singular-mode fRG by Wang \textit{et al.}~\cite{Wang2012}, the so-called `truncated unity fRG' was set up \cite{Lichtenstein2017}. 
This formalism combines various technical improvements and allows for the development of a highly parallelizable and fast-performing code, mainly involving 2D- (or 3D-) integrations and matrix multiplications \cite{ROHE2016160}. The name `truncated unity' comes from the insertion of unity into loop diagrams. These unities are sums over form factors, which are truncated by only considering the relevant form factors. As the form factors can be related to fermion bilinears on the real-space lattice, one obtains a physically appealing understanding what the given truncation captures and what is left out. Furthermore, one can check the convergence of the truncation by varying the number of form factors kept \cite{Lichtenstein2017}. Notably, the truncated unity fRG methodology is useful in another diagrammatic approach, the parquet scheme, whose performance can be boosted significantly in the truncated unity PA \cite{Eckhardt18,Eckhardt2019}. A major step forward was achieved by including the frequency dependence and self-energy corrections into the truncated unity fRG framework \cite{TagliaviniHille2019} and related schemes  \cite{Vilardi2017,Vilardi2019}. Supplemented by multiloop corrections \cite{TagliaviniHille2019}, this paved the way for {\em a)} internal convergence of the fRG as a function of the number of loops, and {\em b)} internal consistency in the fRG, as different ways to compute response functions (by the flow of external-field couplings or by post-processing of the final interaction vertex) and flows with different cutoff schemes are found to agree.

In this paper we now aim at demonstrating the accuracy of our approach to describe one of the most challenging quantum many-body models of correlated electrons: the 2D Hubbard model. We do this by comparing data obtained by the truncated unity fRG in the multiloop extension (hereafter denoted by the fRG* with the adapted self-energy flow and by the fRG with the conventional self-energy flow) with other numerical methods. 
Going beyond the preliminary results of Ref.~\onlinecite{TagliaviniHille2019}, restricted to the calculation of static response functions, we address here the computation of the self-energy in detail. In particular, investigating the form of the multiloop flow equation for the self-energy, we find that a form-factor expansion of the two-particle vertex prevents the full reconstruction of the Schwinger--Dyson equation (SDE) \cite{Schwinger,Dyson} at loop convergence. The deviations can be traced back to the approximations introduced by the form-factor projections in the different channels and, more importantly, can be cured by using the direct derivative of the SDE instead. Including this methodological improvement, we provide a detailed analysis of the quantitative reliability of the fRG for the 2D Hubbard model by verifying the agreement with the solution of the PA. Further, in a comparison to quantum Monte Carlo data, we show that the fRG is remarkably accurate up to moderate interaction strengths. 
This demonstration may have considerable impact as it shows that the {\it fRG can be pushed to become a quantitative quantum many-body method}. One of its benefits for condensed matter material research is the unbiased treatment of various fluctuation channels down to low energy scales, including the indication of ordering transitions and (pseudo-) gap openings. This important feature is accompanied by the conceptual transparency of the approach that can be exploited to pin down the processes responsible for emerging physical effects.
From a longer-term perspective, a quantitatively reliable implementation of the fRG approach provides the most suitable setup for its combination \cite{Taranto2014,Wentzell2015} with complementary quantum many-body approaches, such as the dynamical mean-field theory (DMFT). The improvements described in this paper could thus represent an essential step for an accurate description of two-dimensional electron systems even in the intermediate- to strong-coupling regime. 

The paper is organized as follows: In Section~\ref{sec:2DHubbard} we introduce the Hubbard model and the main observables. In Section~\ref{sec:mfRG} we discuss the multiloop flow equation of the self-energy and present its extension in the truncated unity fRG framework, showing that it correctly accounts for the form-factor projections in the different channels. Section~\ref{sec:numerics} contains a brief description of our benchmarking methods, the PA and determinant quantum Monte Carlo (DQMC).
In Sections~\ref{sec:results} and \ref{sec:results2} we illustrate our results for the self-energy and the susceptibilities for the 2D Hubbard model at half filling and out of it, provide numerical evidence for the convergence to the PA, and perform benchmarks with the DQMC data. 
We conclude with a summary and an outlook in Section~\ref{sec:outlook}.

\section{Model and observables}
\label{sec:2DHubbard}

\subsection{2D Hubbard model}

We consider the single-band Hubbard model in two dimensions,
\begin{align} 
\label{eq:defhamilt}
\hat{\mathcal{H}}=\sum_{i,j, \sigma}t_{ij}\hat{c}^{\dagger}_{i\sigma}\hat{c}_{j\sigma}+U\sum_i \hat{n}_{i\uparrow}\hat{n}_{i\downarrow} - \mu \sum_{i,\sigma}\hat{n}_{i\sigma} \;,
\end{align}
where $\hat{c}_{i\sigma}$ ($\hat{c}^{\dagger}_{i\sigma}$) annihilates (creates) an electron with spin $\sigma$ at the lattice site $i$ ($\hat{n}_{i\sigma}=\hat{c}^{\dagger}_{i\sigma}\hat{c}_{i\sigma}$), $t_{ij}=-t$ is the hopping between nearest-neighbor sites, $t_{ij}=-t'$ is the hopping between next-nearest-neighbor sites, $\mu$ is the chemical potential, and $U$ is the on-site Coulomb interaction. The bare propagator is
\begin{align} \label{eq:G0}
    G_0({\bf k},i\nu)= \big( i\nu +\mu -\epsilon_{\bf k} \big)^{-1} \,,
\end{align}
with
\begin{align}
    \epsilon_{\bf k}=-2t( \cos{k_x} + \cos{k_y} ) - 4t'\cos{k_x}\cos{k_y} \;.
\end{align}
In the following we use $t\equiv 1$ as the energy unit.

\subsection{Susceptibilities and self-energy}

We compute different susceptibilities describing the linear response of a system to a weak external perturbation, as obtained from the fRG*, the PA, and DQMC.
In Matsubara frequency space, the susceptibilities are defined via Fourier transform with respect to imaginary time $\tau$,  
\begin{align}
\label{eq:chieta}
\chi_{\eta}(\mathbf{q},i\omega)
 &= \int_{0}^{\beta} \mathrm{d}\tau \, e^{i\omega\tau} \chi_{\eta}(\mathbf{q},\tau) \;,
\end{align}
where $\eta={\mathrm{M}/\mathrm{D}/\mathrm{SC}}$ indicates the magnetic, density, and superconducting ($s$- and $d$-wave) channels, respectively.

In the half-filled Hubbard model the dominant susceptibility is the antiferromagnetic (AF) one, defined by $\chi_{\mathrm{AF}}=\chi_{\mathrm{M}}({\bf q}=(\pi,\pi),i\omega=0)$ through the magnetic (or spin) susceptibility
\begin{align}
\label{eq:chiM}
\chi_{\mathrm{M}}(\mathbf{q},\tau)
 &= 
\langle\mathit{T}_\tau\hat{s}^z(\mathbf{q},\tau)\hat{s}^z(\mathbf{q},0)\rangle-\langle\hat{s}^z(\mathbf{q},\tau)\rangle \langle\hat{s}^z(\mathbf{q},0)\rangle \;,
\end{align}
where we use the spin operator in the $z$ direction $\hat{s}^z(\mathbf{q},\tau)=\left[\hat{n}_{\uparrow}(\mathbf{q},\tau)-\hat{n}_{\downarrow}(\mathbf{q},\tau)\right]/2$ and the spin-resolved density operator $\hat{n}_{\sigma}(\mathbf{q},\tau)= \sum_{\bf k} \hat{c}_{\sigma}^{\dag}({\bf k}+{\bf q},\tau)\,\hat{c}_{\sigma}({\bf k},\tau)$. 
The sum over momenta includes the normalization factor associated with the momentum integration (or summation) over the first Brillouin zone. 

The density (or charge) response function is defined by
\begin{align}
\label{eq:chiD} 
\chi_{\mathrm{D}}(\mathbf{q},\tau)  &= \tfrac{1}{4}  \left(\langle \mathit{T}_{\tau} \hat{n}(\mathbf{q},\tau)\hat{n}(\mathbf{q},0)\rangle-\langle\hat{n}(\mathbf{q},\tau)\rangle\langle\hat{n}(\mathbf{q},0)\rangle \right)\;,
\end{align}
with $\hat{n}(\mathbf{q},\tau)=\hat{n}_{\uparrow}(\mathbf{q},\tau)+\hat{n}_{\downarrow}(\mathbf{q},\tau)$.
In particular, we will show results for the charge compressibility $\kappa=4\chi_{\mathrm{D}}( {\bf q}=(0,0),i\omega=0)$ and the charge density wave susceptibility $\chi_{\mathrm{CDW}}=\chi_{\mathrm{D}}({\bf q}=(\pi,\pi),i\omega=0)$. 

For the $n=s,d$ pairing susceptibility
\begin{align}
\label{eq:chiSCs}
    \chi_{\mathrm{SC},n}(\mathbf{q},\tau)= \hspace{5cm}\nonumber  \\ \frac{1}{4}\Big\langle\mathit{T}_{\tau}\left[\hat{\Delta}_{n}^{\dag}(\mathbf{q},\tau)+\hat{\Delta}_{n}(\mathbf{q},\tau)\right]\left[\hat{\Delta}_{n}^{\dag}(\mathbf{q},0)+\hat{\Delta}_{n}(\mathbf{q},0)\right]\Big\rangle \;,
\end{align}
we consider both the local $s$-wave pairing
$\hat{\Delta}_s(\mathbf{q},\tau) =  \sum_{\bf k} \hat{c}_{\uparrow}^{\dag}({\bf q}-{\bf k},\tau)\hat{c}_{\downarrow}^{\dag}({\bf k},\tau)$
and the nearest-neighbor $d$-wave pairing
$\hat{\Delta}_d(\mathbf{q},\tau) =  \sum_{\bf k} \left[ \cos(k_x)-\cos(k_y)\right] \hat{c}_{\uparrow}^{\dag}({\bf q}-{\bf k},\tau)\hat{c}_{\downarrow}^{\dag}({\bf k},\tau)$. 
We will focus on the ${\bf q}=(0,0)$ and $i\omega=0$ components, referred to as $\chi_{\mathrm{SC},s}$ and $\chi_{\mathrm{SC},d}$, repsectively.

Furthermore, we compute the self-energy $\Sigma({\bf k},i\nu)=G_0^{-1}({\bf k},i\nu)-G^{-1}({\bf k},i\nu)$, where $G$ is the renormalized propagator $G({\bf k},\tau)=-\langle\mathit{T}_\tau\hat{c}_{\sigma}({\bf k},\tau)\hat{c}_{\sigma}^{\dag}({\bf k},0)\rangle$.
We also show results for the double occupancy (DOC), which can be obtained either from the susceptibilities 
\begin{equation}
\label{eq:doc2p}
    \mathrm{DOC}^{(\mathrm{2P})}=\sum_{i\omega} \int \mathrm{d}{\bf q} \, \left[\chi_{\mathrm{D}}({\bf q},i\omega)-\chi_{\mathrm{M}}({\bf q},i\omega)\right]  + n_{\uparrow}n_{\downarrow} \;
\end{equation}
at the two-particle level or, equivalently, from
\begin{equation}
\label{eq:doc1p}
    \mathrm{DOC}^{(\mathrm{1P})}=\frac{1}{U}\sum_{i\nu} e^{i\nu 0^+} \int \mathrm{d}{\bf k} \, \Sigma({\bf k},i\nu) G ({\bf k},i\nu)
\end{equation}
at the one-particle level.
The sum over frequencies includes the normalization factor of temperature $T$.

\section{Multiloop fRG}
\label{sec:mfRG}

\subsection{Channel decomposition of the vertex}
\label{ssec:vertex_decomposition}
Next to the self-energy, a central object of fRG and parquet algorithms is the (one-particle irreducible) two-particle vertex $F$. 
Using the $SU(2)$ spin symmetry \cite{Rohringer2012}, we can restrict ourselves to one spin component, $V=F^{\uparrow\uparrow\downarrow\downarrow}$.
From the vertex, the susceptibilities can be computed by contracting the two-particle vertex at the end of the flow (``post-processed'') or alternatively via the flow of the response vertices \cite{TagliaviniHille2019} (see Appendix~\ref{app:flowvspp} for a more detailed discussion).

In the channel or parquet decomposition of the vertex we can identify the two-particle reducible contributions $\Phi_{pp/ph/\overline{ph}}$ in the particle-particle, particle-hole and crossed (or transverse) particle-hole channels, respectively. We have
\begingroup
\allowdisplaybreaks[0]
\begin{align} 
    V(k_1,k_2,k_3,k_4)&=\Lambda_{\mathrm{2PI}}+\Phi_{ph}(k_2-k_1,k_1,k_4) \nonumber \\
    &\hspace{-1.7cm}+\Phi_{\overline{ph}}(k_3-k_2,k_1,k_2)+\Phi_{pp}(k_1+k_3,k_1,k_4)\;,
    \label{eq:parquet_decomposition}
\end{align}
\endgroup
where the reducible vertices on the right-hand side (r.h.s.\ ) are parametrized according to a single generalized transfer momentum and two fermionic momenta. In the parquet approximation, the fully two-particle irreducible vertex is approximated by $\Lambda_{\mathrm{2PI}}=U$.
Fully accounting for the interplay between different two-particle channels is a central motivation for the multiloop extension of the fRG, described next.

\subsection{Multiloop extension of fRG: a brief overview}
\label{ssec:overview}

The Wetterich equation is an exact one-loop ($1\ell$) flow equation for the generating functional of one-particle irreducible vertices \cite{Wetterich1993}. Expanding in the vertices leads to an infinite hierarchy of one-loop flow equations for the vertices. However, objects like the three-particle vertex are intractable for numerical treatments. The fundamental approximation in many fRG flows is therefore the truncation in the hierarchy of flow equations \cite{Metzner2012}.
Setting the three-particle vertex to zero yields an approximate $1\ell$ flow equation for the self-energy and two-particle vertex.
A common way to reintroduce some of the lost contributions is to reuse the scale derivative of the self-energy $\dot{\Sigma}$ in the flow of the two-particle vertex. This ``Katanin substitution'' \cite{Katanin2004} already leads to significantly improved results, labeled by $1\ell_K$ in the following. A further refinement, which effectively incorporates the three-particle vertex to third order in the renormalized interaction, is realized by reusing the $1\ell$ results in a $2\ell$ addition to the vertex flow \cite{Katanin2004,Eberlein2014,Rueck2018}. 

The multiloop fRG \cite{Kugler2018a,Kugler2018b} extends these schemes to arbitrary loop order. One finds that the multiloop additions complete the scale derivative of all the diagrams which are only partly generated in the $1\ell$ flow. Thus, they remove the dependence of the results from the particular choice of the regulator; the quantitative reliability of these results is the subject of the present paper. Due to the iterative structure of the multiloop corrections, the numerical effort grows linearly with the loop order. Starting from an efficient algorithm for the $1\ell$ flow, the implementation of the multiloop equations is straightforward.

An alternative derivation of the multiloop flow equations \cite{Kugler2018c} highlights the close connection between fRG and the parquet approach (see also Section~\ref{ssec:Postprocessing}). Starting from the parquet equations, a scale dependence of propagators and vertices can be introduced by making the bare propagator scale dependent. Taking the scale derivative of the self-consistent parquet equations then leads to the multiloop flow equations. The equivalence is exact, provided the self-energy and vertices are treated without further approximation of their momentum and frequency dependence. 

A crucial ingredient for such an equivalence and overall quantitative accuracy is a good resolution of the two-particle vertex and the self-energy in momentum and frequency space. 
Since the numerical effort grows very fast with the number of momentum patches and the size of the frequency window, an efficient vertex parametrization of the two-particle vertex is crucial. 
We use the truncated unity fRG \cite{Maier2013,Lichtenstein2017} for the momentum dependence, together with an accurate treatment of the frequency dependence which includes the high-frequency asymptotics \cite{Wentzell2016}.
For a detailed description and their implementation in the multiloop extension we refer to Refs.~\onlinecite{TagliaviniHille2019,HilleThesis}, where we also provide the expression of the employed smooth frequency cutoff.
Here we further use a refined momentum grid to resolve the peak at $\bf{q}=(\pi,\pi)$ in the AF susceptibility at half filling.
This can be easily accounted for in the truncated unity formulation with precalculated projection matrices in real space \cite{TagliaviniHille2019}.
The accurate description of the long-range AF fluctuations is also of major importance in order to fulfill the Mermin-Wagner theorem \cite{Mermin1966}. 

\begin{figure*}[ht]
    \centering
    \includegraphics[width=0.85\cw]{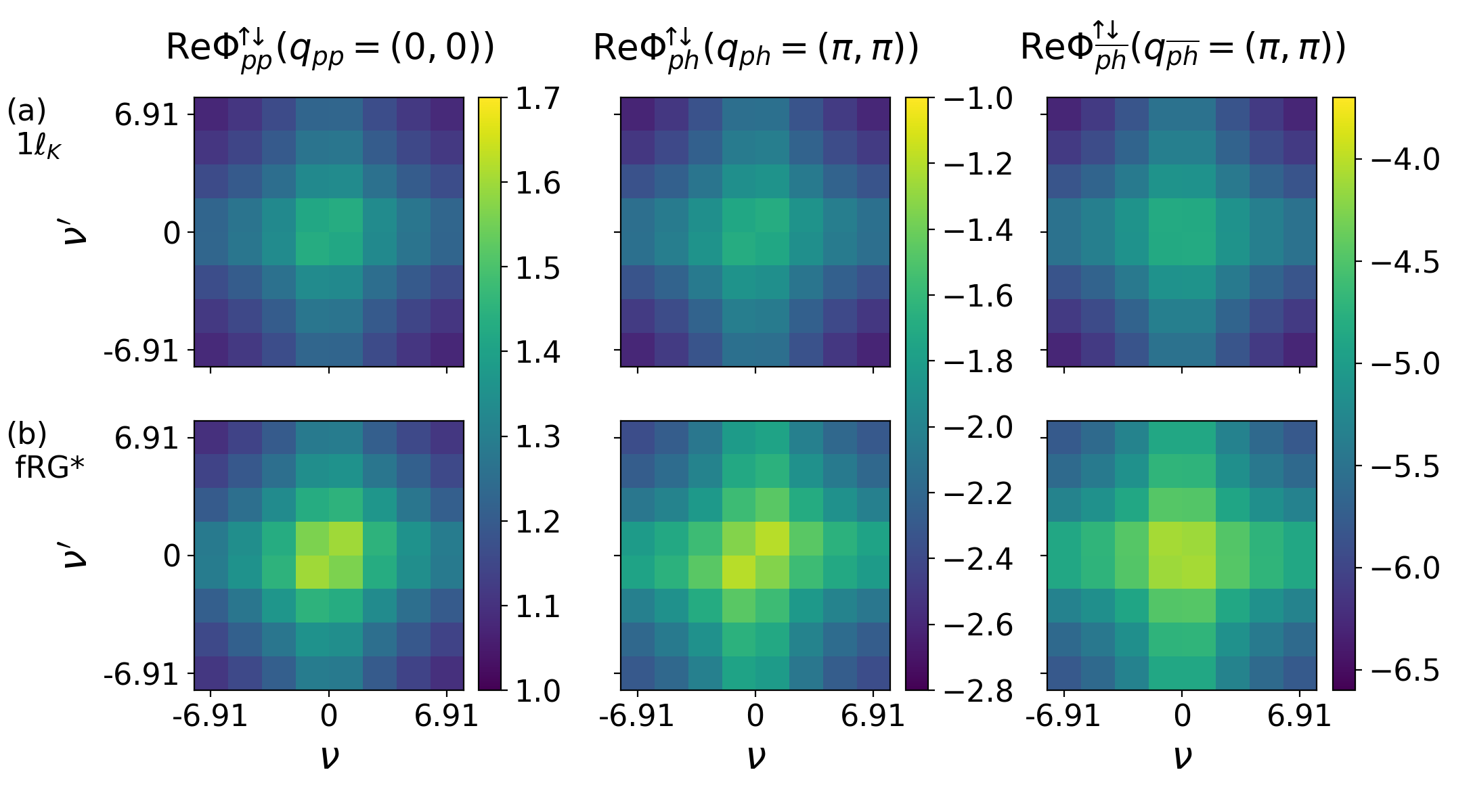}
    \caption{Two-particle vertex decomposed as in Eq.~\eqref{eq:parquet_decomposition} in the different $s$-wave channel contributions at zero bosonic frequency as a function of fermionic frequencies, (a) $1\ell$ including the Katanin correction vs.\ (b) multiloop data, for $U=2$ and $1/T=5$. 
    }
    \label{fig:phi_Kat_SDE_nu_nup}
\end{figure*}

\begin{figure}[hb]
\centering
\includegraphics[width=\cw]{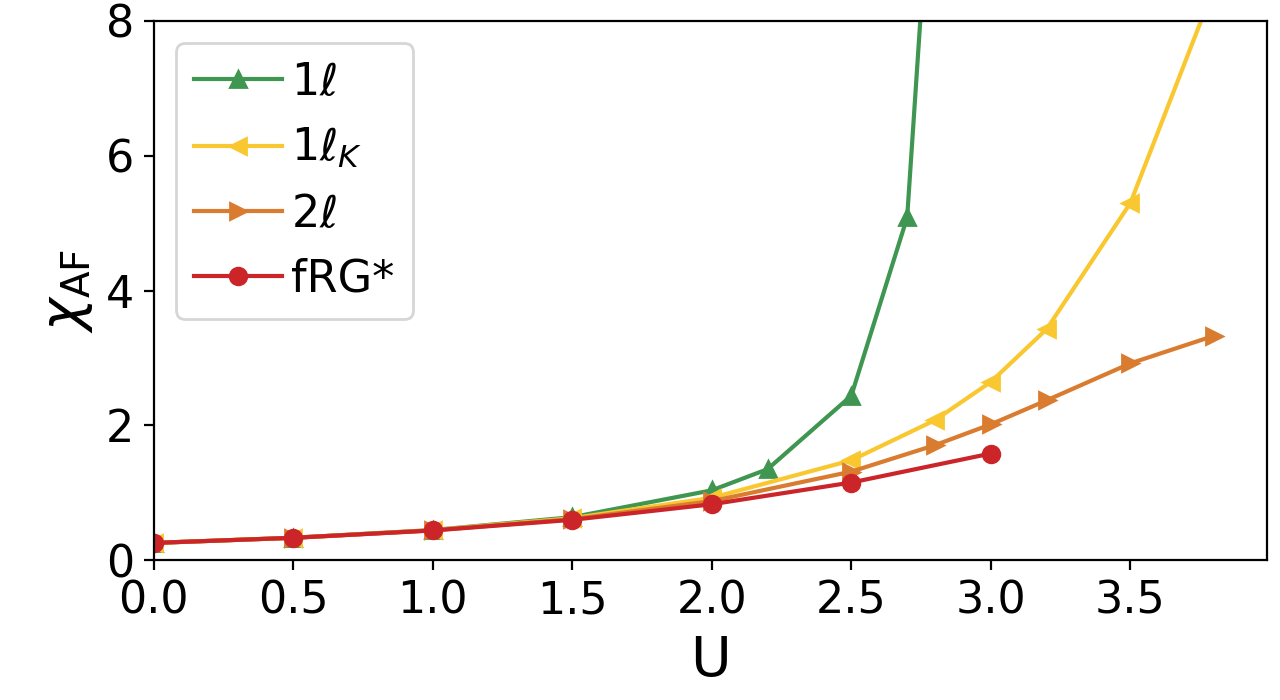}
\caption{Antiferromagnetic susceptibility $\chi_{\mathrm{AF}}(i\omega=0)$ defined in Eq.~\eqref{eq:chiM} as a function of the bare interaction $U$, for $1/T=5$.}
\label{fig:chiAF_loops_B5_scanU}
\end{figure}

The effect of the multiloop extension on the frequency structure of the two-particle vertex $V$ is exemplified in Fig.~\ref{fig:phi_Kat_SDE_nu_nup}.
Here we compare $1\ell_K$ results with fully converged fRG* results, where the latter are obtained by using the revised multiloop flow equation of the self-energy presented in Section \ref{ssec:mfRG_SDE}. 
While $\Phi_{pp}$ is enhanced in the fRG*, $\Phi_{ph}$ and $\Phi_{\overline{ph}}$ are screened, and their absolute values are smaller. The screening of $\Phi_{\overline{ph}}$ with more loops is also reflected in the AF susceptibility, shown in Fig.~\ref{fig:chiAF_loops_B5_scanU}, which is dominated by the contribution of the magnetic channel $\Phi_{M}=-\Phi_{\overline{ph}}$.  
The $1\ell$ scheme strongly overestimates the peak at momentum transfer ${\bf q}=(\pi,\pi)$ leading to an AF ordering at finite interaction strength in violation of the Mermin--Wagner theorem \cite{Mermin1966}. With increasing loop order, the AF peak is reduced. 
At an inverse temperature of $1/T=5$, the $2\ell$ result is already very close to the fRG* on the scale of this plot. 
We note that for $U=3$ the fRG* is not fully converged with respect to frequencies and loop order, and for this reason no results for larger values of $U$ are displayed. The convergence threshold we use is $1\%$ for $\chi_{\mathrm{AF}}$ and for $\mathrm{Im}\,\Sigma$ at momenta $(\pi,0)$ and $(\pi/2,\pi/2)$ and the first two Matsubara frequencies.
For completeness, we report the parameters used for the benchmark analysis of Sections~\ref{sec:results} and~\ref{sec:results2} in Appendix~\ref{app:param}.

\subsection{Parquet approximation and post-processing of the self-energy}
\label{ssec:Postprocessing}

Generally, in many-body theory, there are various exact relations between one- and two-particle quantities, such as the Bethe--Salpeter equations \cite{Bickers2004,Roulet1969} connecting different parts of the two-particle vertex and the SDE relating the self-energy and the vertex. While these equations are used in parquet approaches to iteratively find a self-consistent solution on the one- and two-particle level, applying them to the final self-energy and vertex of any method is a way to check its consistency. 

\begin{figure}[hb]
\centering
	\includegraphics[scale=0.4]{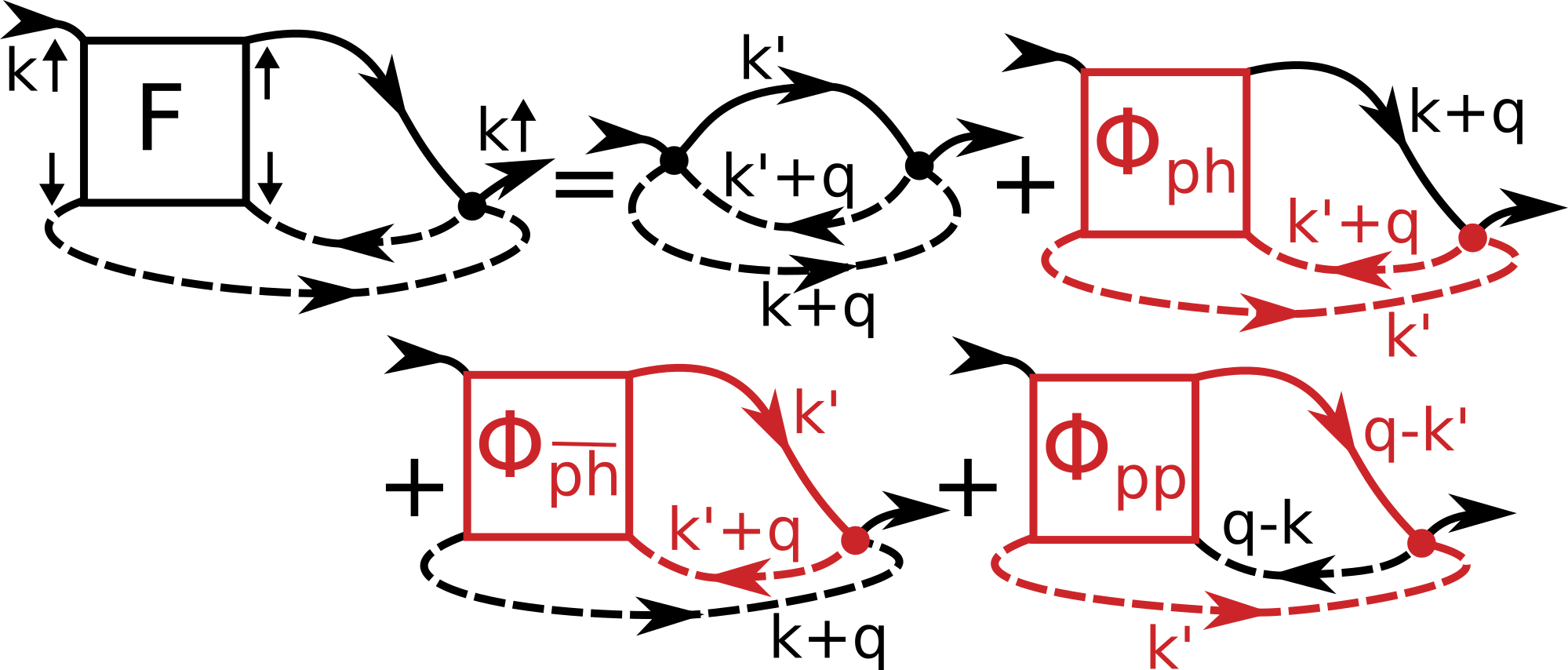}
	\caption{Right-hand side of the Schwinger--Dyson equation for the self-energy \eqref{eq:SDE_channel_decomposed}, illustrating the contributions of the different channels, where  solid  (dashed)  lines  carry  spin up  (down). The first diagram on the r.h.s.\ can be calculated using the convolution theorem and Fast-Fourier-transform algorithms. The other contributions can be determined by first combining the two-particle reducible vertices ($\Phi$) and the bare vertex (dot) through the propagator pair of the corresponding channel (red). Finally, each diagram is closed by a propagator (black) through the direct summation over frequency and momentum.}
	\label{fig:SDE_channels}
\end{figure}

In our previous work \cite{TagliaviniHille2019}, we focused on two-particle quantities like the susceptibilities and compared their outcome directly from the fRG flow with their post-processed result using the final self-energy and vertex.
Here we use the SDE to analogously determine the self-energy.
The SDE involves the self-energy itself through the full propagator as well as the vertex and reads
\begingroup
\allowdisplaybreaks[0]
\begin{align}
    \Sigma({\bf k},i\nu)&=-\sum_{{\bf k' q}} \sum_{i\nu' i\omega} 
    V({\bf k},{\bf k'},{\bf k'}+{\bf q},i\nu,i\nu',i\nu'+i\omega) \nonumber\\
    &\hspace{-.8cm}\times G({\bf k'},i\nu') 
    G({\bf k'}+{\bf q},i\nu'+i\omega)
    G({\bf k}+{\bf q},i\nu+i\omega) \, U \; .
\label{eq:SDE_standard}
\end{align}
\endgroup
Note that the fourth dependence of the vertex is determined by momentum and frequency conservation. Its diagrammatic representation is shown on the left-hand side of Fig.~\ref{fig:SDE_channels}. 
Note that we take the Hartree part implicitly into account by shifting the chemical potential by $U\langle \hat{n}_\sigma \rangle$; half filling then corresponds to $\mu=0$.

In the parquet decomposition~\eqref{eq:parquet_decomposition}, the vertex is split into the fully irreducible part and the three two-particle channels.
As depicted on the right-hand side of Fig. \ref{fig:SDE_channels}, the SDE is then determined by four parts
\begin{widetext}
\begin{align} 
    \Sigma({\bf k},i\nu) &= - \sum_{{\bf k' q}} \sum_{i\nu' i\omega} U^2 G({\bf k'},i\nu')  
   G({\bf k'}+{\bf q},i\nu'+i\omega) G({\bf k}+{\bf q},i\nu+i\omega) \nonumber \\
    & + \sum_{{\bf q}\,i\omega} \sum_{m}f^*_m({\bf k}) \Big[  \sum_{i\nu'} \sum_{n}
    \Phi_{ph,m\,n}({\bf q},i\omega,i\nu,i\nu') 
    \Pi_{ph,n\,0}({\bf q},i\omega,i\nu')
    4\pi^2 U f_{0}({\bf k}) \Big] G({\bf k}+{\bf q},i\nu+i\omega) \nonumber \\
    & + \sum_{{\bf q}\,i\omega} \sum_{m}f^*_m({\bf k}) \Big[  \sum_{i\nu'}\sum_{n} \Phi_{\overline{ph},m\,n}({\bf q},i\omega,i\nu,i\nu')
    \Pi_{ph,n\,0}({\bf q},i\omega,i\nu') 
    4\pi^2 U f_{0}({\bf k}) \Big] G({\bf k}+{\bf q},i\nu+i\omega) \nonumber \\
    & - \sum_{{\bf q}\,i\omega} \sum_{m}f^*_m({\bf k}) \Big[  \sum_{i\nu'} \sum_{n}
    \Phi_{pp,m\,n}({\bf q},i\omega,i\nu,i\nu') 
    \Pi_{pp,n\,0}({\bf q},i\omega,i\nu') 
    4\pi^2 U f_{0}({\bf k}) \Big] G({\bf q}-{\bf k},i\omega-i\nu)\;,
    \label{eq:SDE_channel_decomposed}
\end{align}
\end{widetext}
where $f_{n}({\bf k})$ are the form factors in the truncated unity fRG. The first diagram can be calculated in real space using the convolution theorem twice.
The remaining parts in the $ph$-, $\overline{ph}$-, and $pp$- channels involve
\begin{subequations}
\begin{align}
    \Pi_{ph,nm}({\bf q},i\omega,i\nu)&= -\int \mathrm{d}{\bf p} \, f^*_n({\bf p}) f_{m}({\bf p}) G({\bf p},i\nu) \nonumber \\
    & \hspace{0.7cm} \times G({\bf q}+ {\bf p},i\omega+i\nu) \\
    \Pi_{pp,nm}({\bf q},i\omega,i\nu)&= \int \mathrm{d}{\bf p} \, f^*_n({\bf p}) f_{m}({\bf p}) G({\bf p},i\nu) \nonumber \\
    & \hspace{0.7cm}  \times G({\bf q} - {\bf p},i\omega - i\nu) \;,
\end{align}
\end{subequations}
where we used the same conventions as in Ref.~\onlinecite{TagliaviniHille2019}. 

Upon applying Eq.~\eqref{eq:SDE_channel_decomposed} as a post-processing procedure to compute the self-energy at the end of the fRG flow, we find deviations of up to $20\%$ in both the real and the imaginary part with respect to the solution of the conventional flow equation
\begin{align}
\label{eq:Sigma_1l}
    \dot{\Sigma}^{\Lambda}(k)&=\sum_{k'} \left[(2 V^{\Lambda}(k,k,k')-V^{\Lambda}(k',k,k) \right] S^{\Lambda}(k') \nonumber \\
    &+ \dot{\Sigma}_{\textrm{mfRG-corr}}\;.
\end{align}
Here the first lines corresponds to the $1\ell$ flow equation and the second to the multiloop corrections \cite{Kugler2018b}.
While at $1\ell$, a difference between the flowing and the post-processed result is not surprising, we expect these differences to vanish in a (loop) converged multiloop fRG solution.
As we will show in the following, the remaining discrepancies originate from the truncated form-factor expansion, for which the flow of the self-energy has to be replaced by the direct derivative of the SDE \footnote{For a related fRG scheme starting from the SDE, see Ref.~\onlinecite{Veschgini}.}. 
\nocite{Veschgini} 

\subsection{Self-energy flow in a form-factor expansion}
\label{ssec:mfRG_SDE}

The different fRG results for the self-energy obtained from the (multiloop) flow and the post-processing via the SDE can be traced back to the truncated unity treatment with a reduced number of form factors \footnote{In the limit of an infinite number of form factors the differences would vanish.}. 
In this case, some of the identities used in the general derivation of the self-energy flow \cite{Kugler2018c} do not hold any more. Consequently, the approximation of the vertex in terms of form factors destroys the equivalence of the flow equation and the SDE.
However, this problem can be overcome by using the direct derivative of the SDE instead.

In more detail, each summand of the SDE contains two vertices and three propagators. 
In Ref.~\onlinecite{Kugler2018c}, multiple transformations which interchange these propagators have been used, amounting to a translation between the different two-particle channel descriptions.
However, in the truncated unity parametrization, channel transformations are only information-loss free in the infinite form factor limit. 
With a finite number of form factors, the invariance of the SDE under the exchange of propagators does not hold anymore. 

Here we propose the fRG* extension which exactly reproduces the SDE for the self-energy in a form-factor expansion of the two-particle vertex. Technically, the flow of the self-energy is replaced by the direct derivative of the SDE (for details see Appendix~\ref{app:impl}), which can be found by first introducing a scale ($\Lambda$) dependence in the SDE and then taking the derivative with respect to $\Lambda$ \cite{Kugler2018c}. 

\begin{figure}[ht]
    \centering
    \includegraphics[scale=0.4]{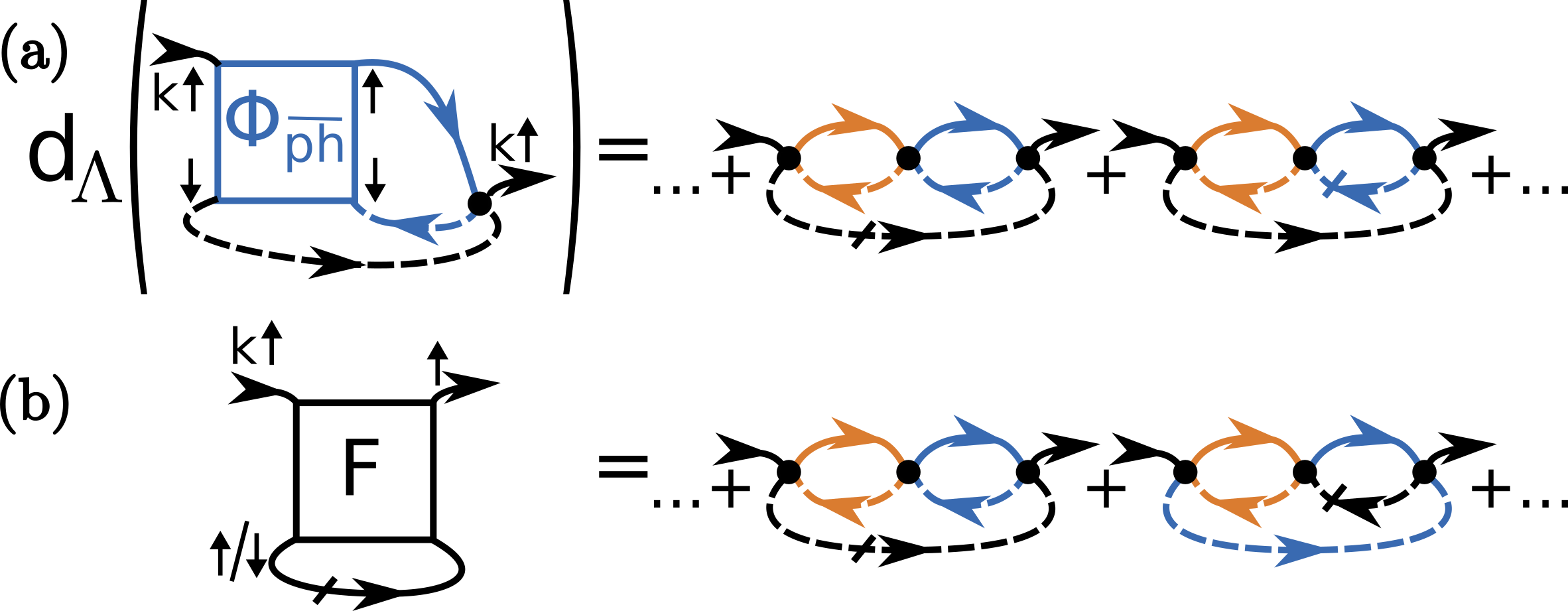}
    \caption{Illustration of the self-energy flows in (a) the fRG*, restricted to the part with $\Phi_{\overline{ph}}$, and (b) the fRG. The r.h.s.\ shows two exemplary differentiated diagrams contributing to the self-energy flow at third order, where solid (dashed) lines carry spin up (down), and the diagonal dash symbolizes a scale-differentiated bare propagator. In the fRG*, two bubbles from the same channel (colored) are combined and then closed with the black line. By contrast, in the fRG, the second diagram requires one to insert a $\overline{ph}$ (orange) into a $pp$ (blue) bubble, before closing with the differentiated propagator (black).}
    \label{fig:Diagrams_SDE_flow}
\end{figure}

An example illustrating the advantage of the new fRG* self-energy flow compared to the conventional flow is shown in Fig.~\ref{fig:Diagrams_SDE_flow}. 
We focus on two specific differentiated diagrams contributing to the flow of the self-energy. 
In the fRG*, each summand of the SDE is directly differentiated with respect to $\Lambda$. 
For concreteness, we consider the penultimate summand on the r.h.s.\ of Fig.~\ref{fig:SDE_channels}, insert the lowest-order diagram for $\Phi_{\overline{ph}}$, and let the $\Lambda$ derivative act on the two spin-down propagators (dashed lines). 
Thereby, we obtain the two differentiated self-energy diagrams on the r.h.s.\ of Fig.~\ref{fig:Diagrams_SDE_flow}~(a), where the propagators with a diagonal dash symbolize differentiated propagators.
The same contributions should be part of the standard fRG self-energy flow, shown in Fig.~\ref{fig:Diagrams_SDE_flow}~(b).
Indeed, the first diagram on the r.h.s.\ simply follows from a second-order $\Phi_{\overline{ph}}$ diagram with two $\overline{ph}$ bubbles, and the second one originates from the $\Phi_{pp}$ part of $F$, where the $\overline{ph}$ bubble (orange) is inserted into another $pp$ bubble (blue).
The crucial point is that the r.h.s.\ of both panels are formally equivalent, but the form-factor truncation applies in a less favorable way on the right diagram in the fRG shown in Fig.~\ref{fig:Diagrams_SDE_flow}: 
After each colored bubble, a truncated unity projects the dependence on the `fermionic' momenta onto a finite number of form factors.
The two diagrams in Fig.~\ref{fig:Diagrams_SDE_flow}~(a) and the first in Fig.~\ref{fig:Diagrams_SDE_flow}~(b) are exactly described by only $s$-wave form factors. 
However, when evaluating the lower right diagram in an $s$-wave form-factor truncation, the $\overline{ph}$ contribution is completely averaged in the process of translating it to the $pp$ channel, thus yielding significantly less accurate results.

\begin{figure}[htb]
    \centering
    \includegraphics[width=\cw]{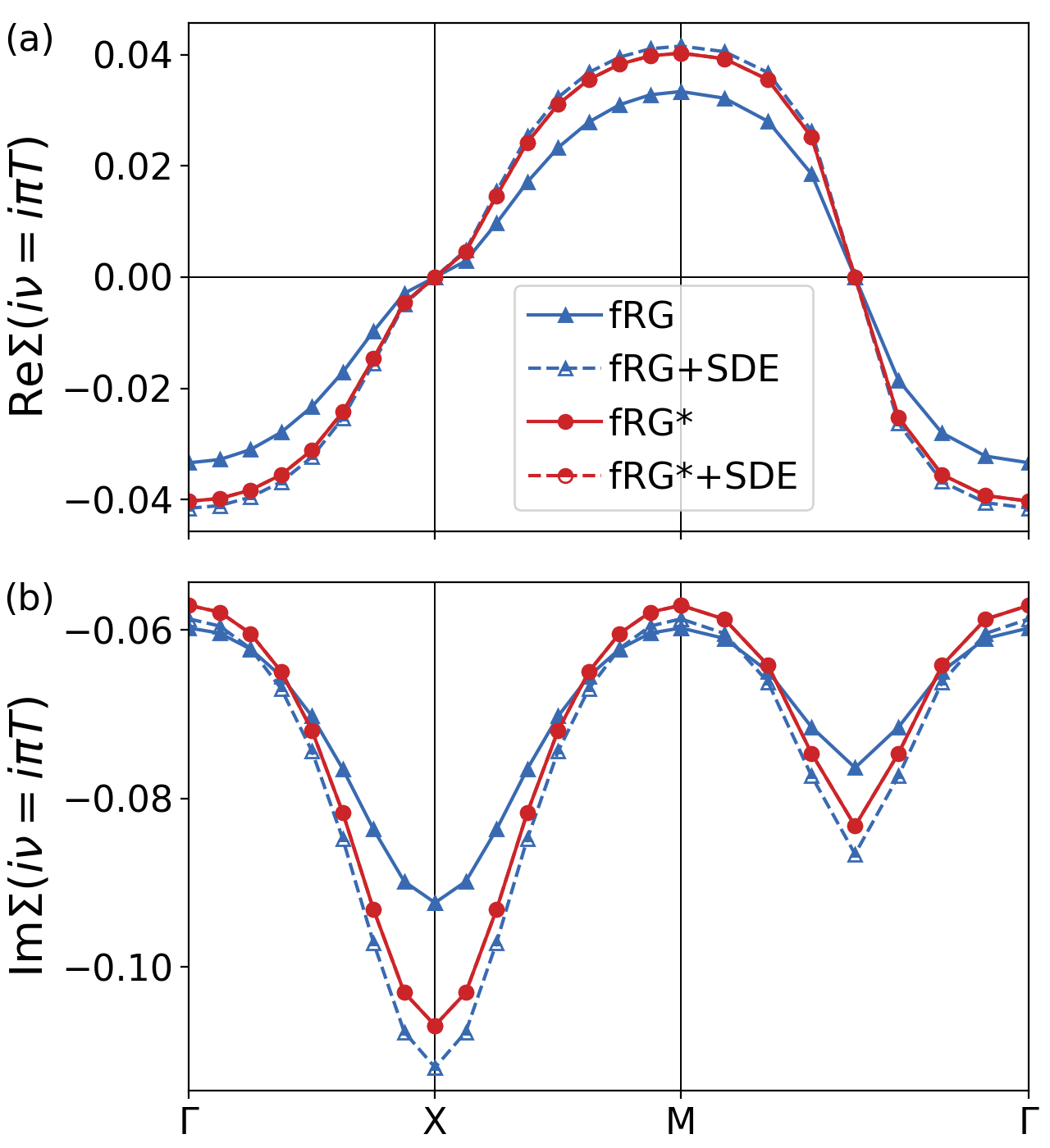}
    \caption{Real (a) and imaginary part (b) of the self-energy as obtained by conventional fRG (blue) and the fRG* flow (red), together with the respective post-processed results (dashed lines), for $U=2$ and $1/T=5$. Within the fRG*, the post-processed (dashed red) ones lie exactly on top of the fRG* flow results (red solid).}
    \label{fig:Sig_flow_SDE_postproc_U2_B5}
\end{figure}

In Fig.~\ref{fig:Sig_flow_SDE_postproc_U2_B5} we present the real and imaginary parts of the self-energy $\Sigma(i\nu=i\pi T)$ as a function of momentum, for $U=2$ and $1/T=5$, both for the conventional fRG (blue) and fRG* (red). We show that, unlike fRG, fRG* yields excellent agreement between the flowing (solid lines) and the post-processed self-energy (dashed lines), determined by Eq.~\eqref{eq:SDE_channel_decomposed} with the final vertices and self-energy at the end of the flow. The non-improved fRG results obtained from the flow exhibit pronounced deviations with respect to the fRG* self-energy. For the post-processed ones, these deviations are significantly reduced since in this case the self-energy is updated by the SDE at the end of the flow.

\subsection{Self-energy iterations}

\begin{figure}[htb]
    \centering
    \includegraphics[width=\cw]{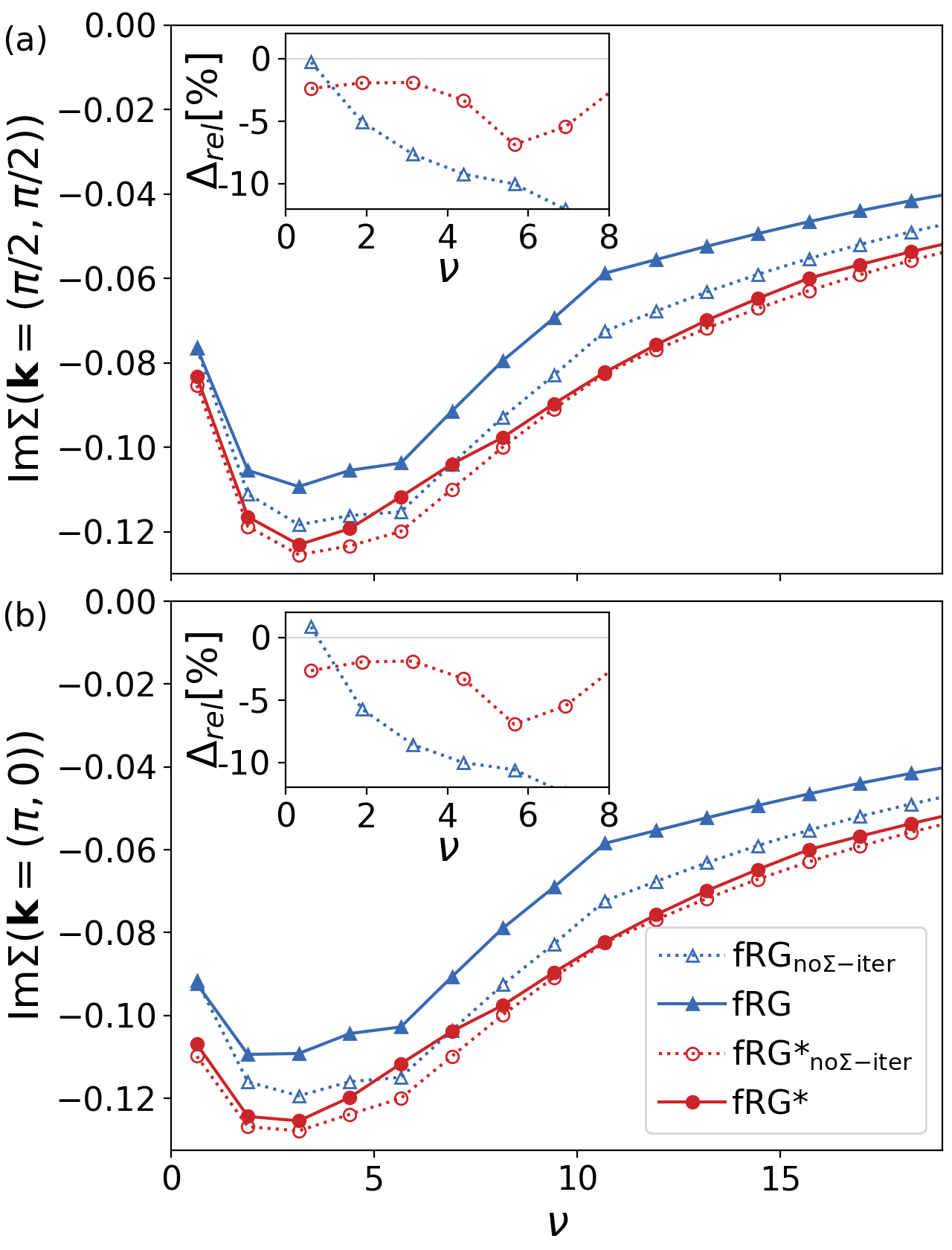}
    \caption{Imaginary part of the self-energy at the (a) nodal and (b) antinodal point, as obtained by the fRG (blue) and fRG*  (red), with and without self-energy iterations, for $U=2$ and $1/T=5$. The inset shows the relative difference caused by neglected self-energy iterations.}
    \label{fig:Sig_n_an_flow_SDE_SEit_U2_B5}
\end{figure}

We now analyze the effect of the self-energy iterations, i.e., a repeated evaluation of the r.h.s.\ of the flow equation, in the fRG as well as in the fRG*. 
As described in detail in Appendix~\ref{app:impl}, these are needed because the r.h.s.\ of both the vertex and the self-energy depend on each other:
In the vertex flow, the Katanin substitution $S({\bf k},i\nu)=\mathrm{d}_{\Lambda}G({\bf k},i\nu)|_{\dot{\Sigma}=0} \rightarrow S^K({\bf k},i\nu)=\mathrm{d}_{\Lambda}G({\bf k}, i\nu)$ depends on the self-energy flow, while the multiloop flow of $\Sigma$ involves the vertex flow, either through the multiloop corrections of Ref.~\onlinecite{Kugler2018b} or through the parts of the differentiated SDE where $\mathrm{d}_\Lambda$ acts on a vertex.
In order to study the effect of the self-energy iterations, we compare the fully converged results to those obtained by solely using the $1\ell$ self-energy flow in the Katanin substitution. 
Figure~\ref{fig:Sig_n_an_flow_SDE_SEit_U2_B5} displays self-energy results with (solid lines) and without (dotted lines) iterations, as obtained by the fRG (blue) and fRG* (red).
We note that the slight kink between the 4$^{\mathrm{th}}$ and 5$^{\mathrm{th}}$ frequency and also between the 8$^{\mathrm{th}}$ and 9$^{\mathrm{th}}$ frequency corresponds to the crossing of the low-frequency tensor range and the high-frequency asymptotics of the two-particle vertex. 
This effect is more pronounced in the fRG than in the fRG*, since the channel-reducible vertices that directly enter the conventional flow \eqref{eq:Sigma_1l} have a richer frequency dependence than those that are first combined with the bare interaction, as needed for the calculation of the SDE-inspired fRG* flow \eqref{eq:SDE_channel_decomposed}.
In the conventional fRG flow, the effect of the self-energy iterations amounts to $5$--$10\%$ except for the first Matsubara frequency, which appears to be well described without any additional iterations.
In contrast, in the fRG*, which accounts for the form-factor projections in the different channels, the self-energy iterations lead to much smaller overall corrections but are relevant for the lowest frequencies (see inset in Fig.~\ref{fig:Sig_n_an_flow_SDE_SEit_U2_B5}).

\begin{figure}[htb]
\centering
\includegraphics[width=\cw]{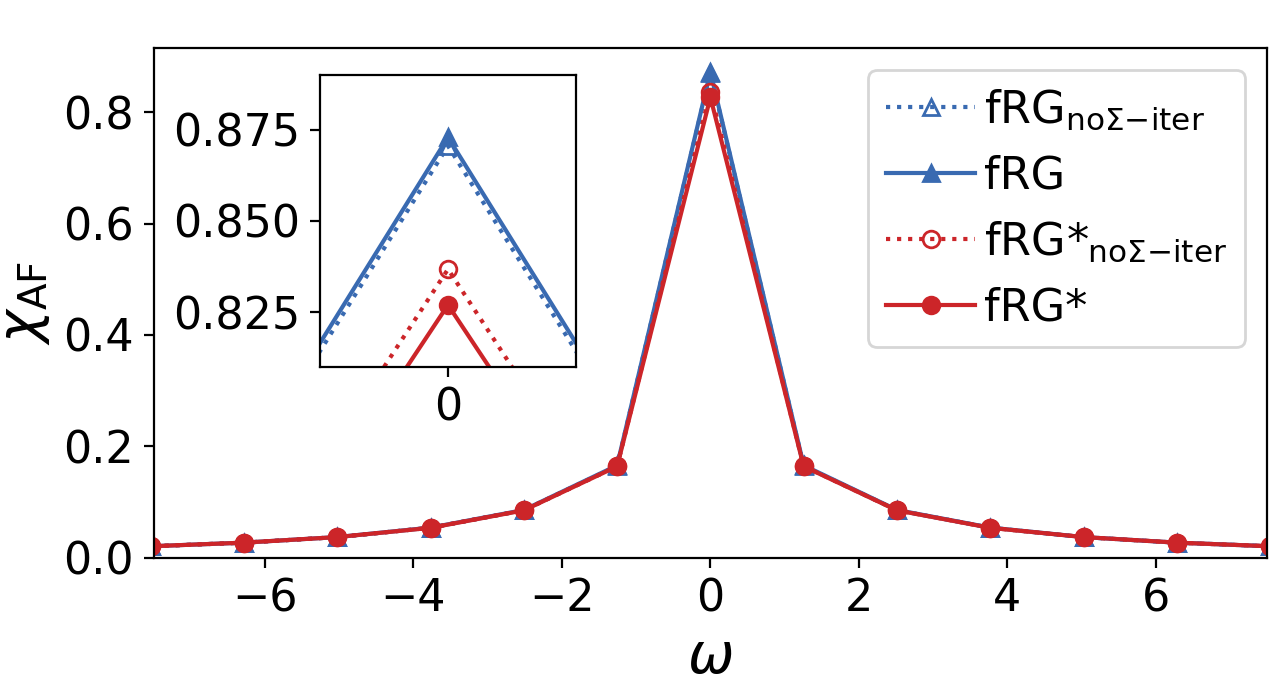}
\caption{Frequency dependence of the AF susceptibility $\chi_{\mathrm{AF}}(i\omega)$ as obtained by conventional fRG (blue) and the fRG* flow (red), with (solid line) and without (dotted line) $\Sigma$ iteration, for $U=2$ and $1/T=5$.}
\label{fig:chiAF_flow_SDE_SEit_U2_B5}
\end{figure}

The self-energy iterations also affect the AF susceptibility $\chi_{\mathrm{AF}}$ displayed in Fig.~\ref{fig:chiAF_flow_SDE_SEit_U2_B5}. For $U=2$ and $1/T=5$, a sizeable effect is only observed for zero Matsubara frequency, highlighted in the inset. 
Neglecting the self-energy iterations in the fRG* (red dotted lines) overestimates the AF peak by $1\%$. In fRG, their impact on $\chi_{\mathrm{AF}}$ is smaller, according to the small effect on the first Matsubara frequency of the self-energy observed in Fig.~\ref{fig:Sig_n_an_flow_SDE_SEit_U2_B5}.
Anticipating the comparison of the fRG* (red) to the PA (gray) (see Fig.~\ref{fig:chiAF_methods_compare}), we remark that, while the AF peak in the conventional scheme (blue) deviates by $5\%$ from the PA, fRG* (red) shows perfect agreement. 

\section{Benchmark methods}
\label{sec:numerics}

\subsection{Parquet approximation}
\label{ssec:parquet}

The PA results are obtained with the truncated unity implementation of the parquet equations \cite{Eckhardt2019}.
The parquet equations are solved by iterating the Bethe--Salpeter equations, the parquet equation \eqref{eq:parquet_decomposition}, and the SDE \eqref{eq:SDE_channel_decomposed} until self-consistency is reached. The momentum dependence of the vertices is parametrized using the form-factor expansion \cite{Eckhardt18}, identical to the scheme used in the truncated unity fRG. Although the equivalence of the PA and the multiloop fRG has been formally shown \cite{Kugler2018c}, the actual PA calculations substantially differ from the ones in the fRG, since no flow parameter is introduced. In the PA no differential equations are solved but the convergence to a fixed point is achieved iteratively, starting from an initial guess for the vertices (in our case given by the lowest-order diagrams).

In order to account for the finite frequency box, we use the asympotics as introduced in Ref.~\onlinecite{Li2016} and also used in Ref.~\onlinecite{Li2019}.
The implementation of the frequency asymptotics thus differs from the one used in the fRG calculations.
All further computational details of the truncated unity implementation of the parquet equations can be found in Ref.~\onlinecite{Eckhardt2019}, as well as a detailed analysis of the convergence in the number of form factors, showing that the approximation of a single form factor, used here at half filling, is justified.

All results are converged in the number of discrete lattice momenta $N_{q}$ and positive fermionic Matsubara frequencies $N_{f+}$.
Specifically, we use a uniform momentum grid with $N_{q} = 32 \times 32$, and a frequency box with $N_{f+} = 32$ positive Matsubara frequencies.

\subsection{Determinant quantum Monte Carlo}
\label{ssec:dQMC}

The DQMC algorithm, proposed by Blankenbecler {\it et al.}~\cite{Blankenbecler1981a}, is a state-of-the-art numerically exact method and is commonly applied for finite-temperature~\cite{Blankenbecler1981a,Blankenbecler1981b} calculations of interacting fermion systems. The basic idea of the DQMC algorithm is to decouple the two-body interaction into noninteracting fermions coupled with auxiliary fields, and to compute the fermionic observables via importance sampling of the fields. To achieve that, a Trotter decomposition and a Hubbard--Stratonovich transformation are successively used, after discretizing the inverse temperature as $\beta=M\Delta\tau$. The systematic error from finite $\Delta\tau$ can be removed by extrapolations with several different $\Delta\tau$ values. For further details, we refer to the reviews~\cite{AssaadEvertz2008,Chang2015}. In this work, we choose $\Delta\tau t=0.02$ which has been tested to safely reach the $\Delta\tau\to0$ limit. In this work, we have also implemented our most recent improvements~\cite{Yuan19a,Yuan19b} of the DQMC algorithm. For the computation of dynamical quantities, we first measure the imaginary-time correlation functions, and then obtain the imaginary-frequency observables via a Fourier transformation. Specifically, for the self-energy we implemented the Legendre polynomial representation~\cite{Boehnke2011} for the imaginary-time single-particle Green's function $G(\mathbf{k},\tau)$ to compute $G(\mathbf{k},i\nu)$, and subsequently applied the Dyson equation. This yields smooth self-energy results even for high frequencies. All DQMC data presented here are found to converge to the thermodynamic limit for a linear system size of $L=28$ (with the number of lattice sites being $N=L^2$) for half filling and $L=24$ away from it. As for statistics, we typically use in total $10^5$ measurement samples after the Markov chain equilibrium process. The error bars are significantly smaller than the corresponding symbol and thus neglected in the plots.

\section{Results at half filling}
\label{sec:results}

We now compare different physically relevant quantities as obtained from fRG*, the PA, and the numerically exact DQMC. In particular, we first focus on the various susceptibilities in Section~\ref{ssec:susc} and then present the results for the self-energy and double occupancy in Sections~\ref{ssec:sigma} and \ref{ssec:doc}, respectively.

\subsection{Susceptibilities}
\label{ssec:susc}

\begin{figure}[tb]
    \centering
    \includegraphics[width=\cw]{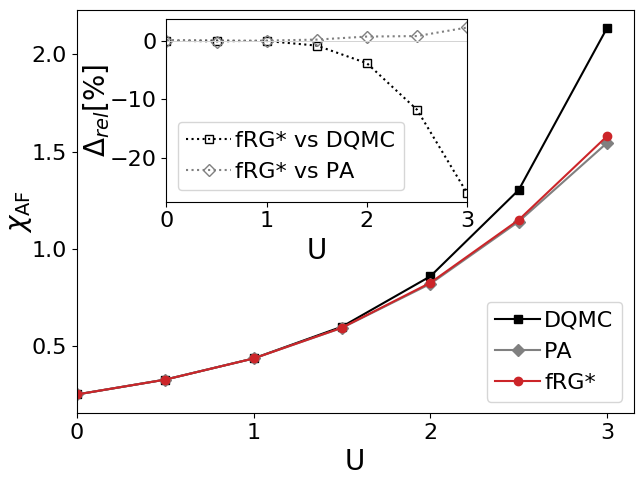}
    \caption{Antiferromagnetic susceptibility $\chi_{\mathrm{AF}}$ as a function of $U$, 
    as obtained by the fRG* (red), the PA (gray), and DQMC (black),
    for $1/T=5$. The inset shows the relative difference.}
    \label{fig:chiAF_methods_compare}
\end{figure}

We first present the results for the leading AF susceptibility, $\chi_{\mathrm{AF}}$, in the half-filled 2D Hubbard model as a function of the bare interaction strength $U$. In Fig.~\ref{fig:chiAF_methods_compare} we report fRG* (red), PA (gray), and DQMC data (black), together with the relative difference of the fRG* with respect to PA and DQMC shown in the inset. Up to $U=2.5$, fRG* and PA coincide with a relative difference of $\leq 1\%$. 
For larger values of $U$, the convergence of the fRG* in frequencies and also in loop numbers becomes numerically challenging and is not reached yet, see also Appendix~\ref{app:flowvspp}.
This leads to the observed deviations from the PA solution. 
The differences between the PA and the numerically exact DQMC data are essentially due to the fully two-particle irreducible diagrams not included in the PA. 
These diagrams contribute to fourth order in $U$; the corresponding relative difference amounts to $\Delta_{\textrm{rel}}\simeq0.05\, U^4$.
A second source of the differences between the PA solution and DQMC is given by the form-factor expansion of the two-particle vertex which accounts only for the local $s$-wave part. Due to perfect nesting, the physics at half filling is dominated by magnetic fluctuations peaked at $\bf{q}=(\pi,\pi)$, and, at small coupling, there are only minor quantitative corrections due to the form-factor truncation \cite{Eckhardt2019}.
Away from half filling, we expect superconducting $d$-wave components to become relevant and hence include those form factors in Section~\ref{sec:results2}, too.

\begin{figure}[tb]
    \centering
    \includegraphics[width=\cw]{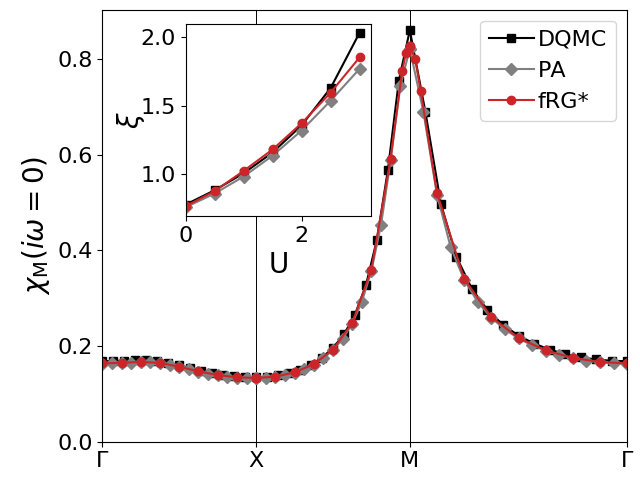}
    \caption{Magnetic susceptibility $\chi_{\mathrm{M}}({\bf q},i\omega=0$) as obtained by the fRG* (red), the PA (gray), and DQMC (black), for $U=2$ and $1/T=5$. The inset shows the correlation length $\xi$ extracted from $\chi_{\mathrm{M}}({\bf q},i\omega=0)$ as a function of $U$, for $1/T=5$ (see the text for details of the fitting procedure).}
    \label{fig:dyn_structureFact_GXMG_methods_compare}
\end{figure}

\begin{figure}
    \centering
    \includegraphics[width=\cw]{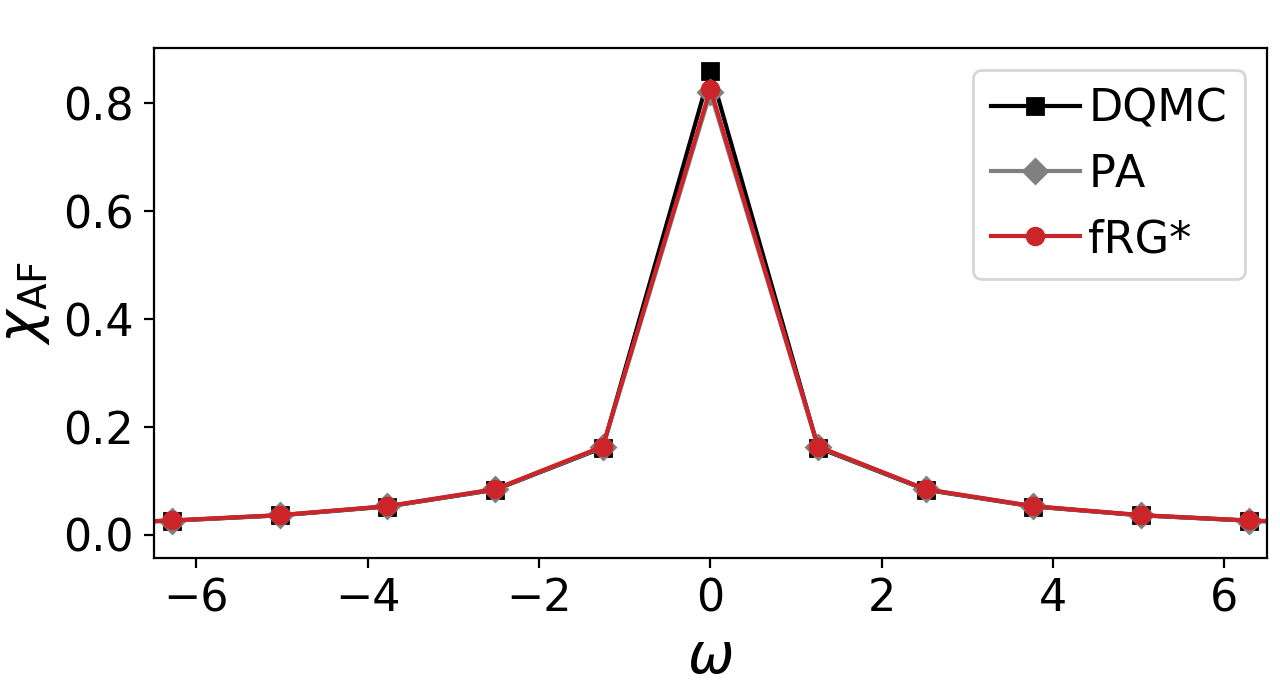}
    \caption{Frequency dependence of the AF susceptibility $\chi_{\mathrm{AF}}(i\omega)$ as obtained by the fRG* (red), the PA (gray), and DQMC (black) for $U=2$ and $1/T=5$.}
    \label{fig:chiAF_iw_methods_compare}
\end{figure}

Figure~\ref{fig:dyn_structureFact_GXMG_methods_compare} shows the momentum dependence of the magnetic susceptibility at zero frequency $\chi_{\mathrm{M}}({\bf q},i\omega=0)$, for $U=2$ (and $1/T=5$).
The results of the fRG*, the PA, and DQMC  exhibit excellent quantitative agreement. The largest deviation is found at $M=(\pi,\pi)$ corresponding to the AF susceptibility shown in Fig.~\ref{fig:chiAF_methods_compare} for different values of $U$.
We note that for all other frequencies $i\omega\neq 0$ the AF susceptibility of the fRG* perfectly agrees with the one of the PA.
While the AF peak height obtained from fRG* does not converge perfectly to the PA for $U>2$, the correlation length $\xi$ extracted from its width shows a very good agreement between the different methods (see the inset in Fig.~\ref{fig:dyn_structureFact_GXMG_methods_compare}. The correlation length is fitted to all points of $\chi_{\mathrm{M}}({\bf q},i\omega=0)$ within a distance of $0.3\pi$ from $M$ through
\begin{align}
    \chi({\bf q},i\omega=0) &\sim \frac{1}{4 \sin^2(\frac{q_x-\pi}{2}) + 4 \sin^2(\frac{q_y-\pi}{2}) + \xi^{-2} }\;,
    \label{eq:corr_length}
\end{align}
which reduces to the Ornstein--Zernike form for small momentum differences $q_x-\pi$ and $q_y-\pi$ (cf.\ Refs.~\onlinecite{Rohringer2011,Schaefer2020}). The number of momenta taken into account for the fit are between $33$ and $45$ in the fRG*, between $69$ and $161$ in PA, and $57$ in DQMC. The maximal standard deviation error is $0.023$ in the fRG*, $0.025$ in PA, and $0.028$ in DQMC. 

The frequency dependence of $\chi_{\mathrm{AF}}$ is shown in Fig.~\ref{fig:chiAF_iw_methods_compare} for $U=2$ and $1/T=5$. The fRG* and PA compare very well at any Matsubara frequency; the largest deviation between DQMC and fRG* is found at zero Matsubara frequency.

\begin{figure}[tb]
    \centering
    \includegraphics[width=\cw]{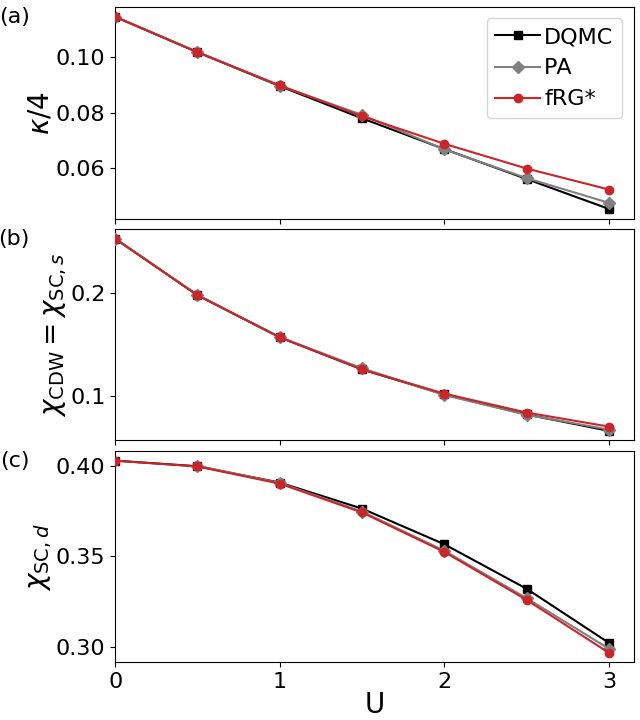}
    \caption{(a) Compressibility $\kappa$, (b) charge density wave $\chi_{\mathrm{CDW}}=\chi_{\mathrm{SC},s}$, and (c) superconducting susceptibility $\chi_{\mathrm{SC},d}(i\omega=0)$ as a function of $U$, as obtained by the fRG* (red), the PA (gray), and DQMC (black), for $1/T=5$.}
    \label{fig:chi_methods_compare}
\end{figure}

In Fig.~\ref{fig:chi_methods_compare} we show different subleading susceptibilities as a function of $U$, for $1/T=5$: 
the compressibility $\kappa$, the ($s$-wave) charge density wave $\chi_{\mathrm{CDW}}$ susceptibility which equals $\chi_{\mathrm{SC},s}$ for $SU(2)$ spin and charge (particle-hole) symmetry (see Appendix \ref{app:CDW_vs_SCs} for the proof), and the $d$-wave superconducting susceptibility $\chi_{\mathrm{SC},d}$.
The quantitative agreement between fRG*, PA, and DQMC results for the subleading susceptibilities is very good, with the relative difference of the fRG* with respect to the PA at $U=3$ being $10\%$ for $\kappa$, $4\%$ for $\chi_{\mathrm{CDW}}$, and less than $1\%$ for $\chi_{\mathrm{SC},d}$, and with respect to DQMC $15\%$ for $\kappa$, $6\%$ for $\chi_{\mathrm{CDW}}$, and $2\%$ for $\chi_{\mathrm{SC},d}$. 
Note that the compressibility $\kappa$ is also consistent with Ref.~\onlinecite{Kim2019}.
The good agreement between fRG* and PA, both affected by the form-factor truncation, and the exact DQMC justifies a computation with only the local $s$-wave form factor.

\begin{figure}[tb]
    \centering
    \includegraphics[width=\cw]{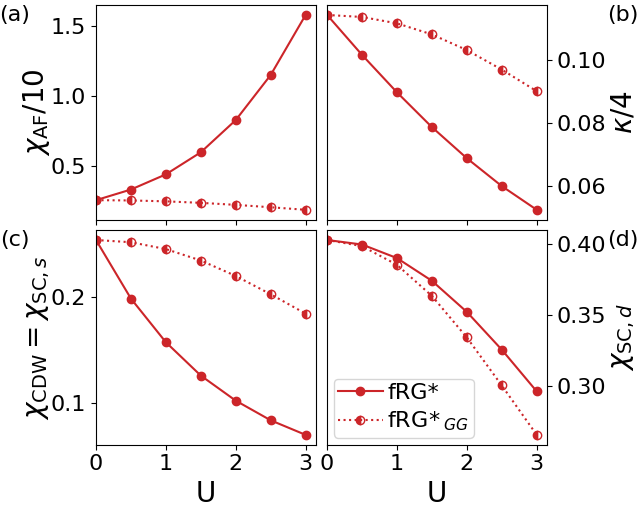}
    \caption{(a) Antiferromagnetic susceptibility $\chi_{\mathrm{AF}}$, (b) compressibility $\kappa$, (c) charge density wave $\chi_{\mathrm{CDW}}=\chi_{\mathrm{SC},s}$ (see Appendix~\ref{app:CDW_vs_SCs}) and (d) $d$-wave superconducting susceptibility $\chi_{\mathrm{SC}}$ as a function of $U$, as obtained by the fRG* with (solid lines) and without (dashed lines) vertex corrections, for $1/T=5$.}
    \label{fig:chi_pp_GG}
\end{figure}

All subleading susceptibilities $\kappa$, $\chi_{\mathrm{CDW}}$, $\chi_{\mathrm{SC},s}$ and $\chi_{\mathrm{SC},d}$ decrease with $U$, since the growing AF fluctuations lead to stronger screening of the subleading fluctuations. Figure~\ref{fig:chi_pp_GG} shows a more detailed analysis of the different contributions to the susceptibility and particularly of the importance of the vertex corrections.
The uncorrelated susceptibilities in terms of dressed Green's functions (without vertex corrections) are determined by
\begin{subequations}
\begin{align}
    \chi_{\mathrm{AF},GG}&= \chi_{\mathrm{CDW},GG}\nonumber \\
    &= \frac{1}{2} \sum_{i\nu}\Pi_{ph,0\,0}({\bf q}=(\pi,\pi),i\omega=0,i\nu)\\
    \kappa_{GG}&=2\sum_{i\nu}\Pi_{ph,0\,0}({\bf q}=(0,0),i\omega=0,i\nu)\\
    \chi_{\mathrm{SC},d,GG}&=\frac{1}{2} \sum_{i\nu}\Pi_{pp,1\,1}({\bf q}=(0,0),i\omega=0,i\nu)\;,
\end{align}
\end{subequations}
where the form-factor index $0$ stands for $s$-wave and $1$ for $d$-wave. They all decrease with $U$, as a consequence of self-energy screening effects. The vertex contributions, given by the difference from the full susceptibilities, exhibit a richer physical behavior: They lead to a reduction or screening of the bare $\kappa$ and $\chi_{\mathrm{SC},s}$ susceptibilities, whereas $\chi_{\mathrm{SC},d}$ and most prominently $\chi_{\mathrm{AF}}$ are enhanced with respect to their bare values. For $\chi_{\mathrm{SC},d}$ the vertex corrections are not strong enough to induce an overall increasing susceptibility. This occurs only for $\chi_{\mathrm{AF}}$, where the vertex corrections are indeed dominant.

\subsection{Self-energy}
\label{ssec:sigma}

\begin{figure}[hb]
    \centering
    \includegraphics[width=\cw]{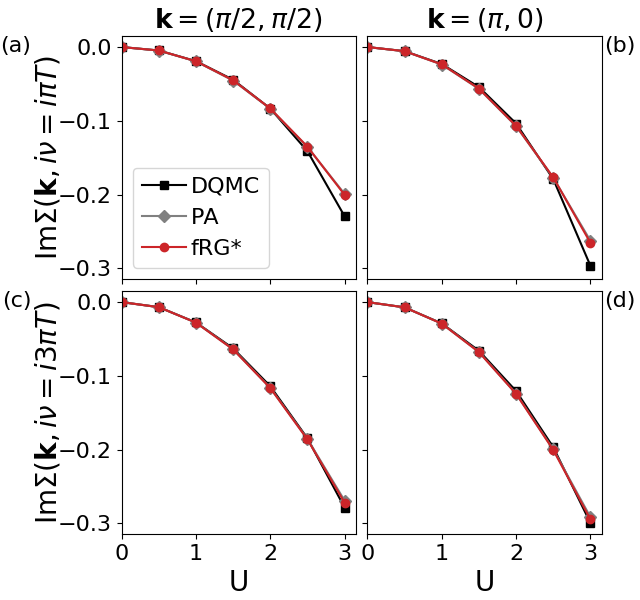}
    \caption{Imaginary part of the self-energy at the (a) first frequency and nodal point, (b) first frequency and antinodal point, (c) second frequency and nodal point, (d) second frequency and antinodal point as a function of $U$, as obtained by the fRG* (red), the PA (gray), and DQMC (black), for $1/T=5$.}
    \label{fig:Sig_methods_compare}
\end{figure}

We now discuss the frequency and momentum dependence of the self-energy and their comparison between the different methods. In Fig.~\ref{fig:Sig_methods_compare} we show the imaginary part at the nodal ${\bf k}=(\pi/2,\pi/2)$ and antinodal ${\bf k}=(\pi,0)$ point as a function of $U$, for $1/T=5$. 
The agreement between fRG* and the PA is almost perfect for small values of $U$, with increasing deviations up to a few percent for larger $U$. However, at $U=3$, the DQMC results for the first Matsubara frequency differ considerably from those of the fRG* and the PA. Moreover, comparing the results for the first and second Matsubara frequency at ${\bf k}=(\pi,0)$ (last data points in Figs.~\ref{fig:Sig_methods_compare}~(b) and~\ref{fig:Sig_methods_compare}~(d)) reveals a discrepancy with an important physical implication: The onset of the pseudogap opening \cite{Schaefer2015b,Schaefer2015c,Simkovic2020}, resulting from quasi-identical values for the first two Matsubara frequencies at ${\bf k}=(\pi,0)$, is observed at $U=3$ in DQMC but not (yet) in the fRG* and the PA. The reason is the neglect of fully two-particle irreducible diagrams. In fRG* and the PA, the absolute value at the first Matsubara frequency is $11\%$ smaller than at the second one, and the gap opening sets in only for larger interactions (not shown).

\begin{figure}[tb]
    \centering
    \includegraphics[width=\cw]{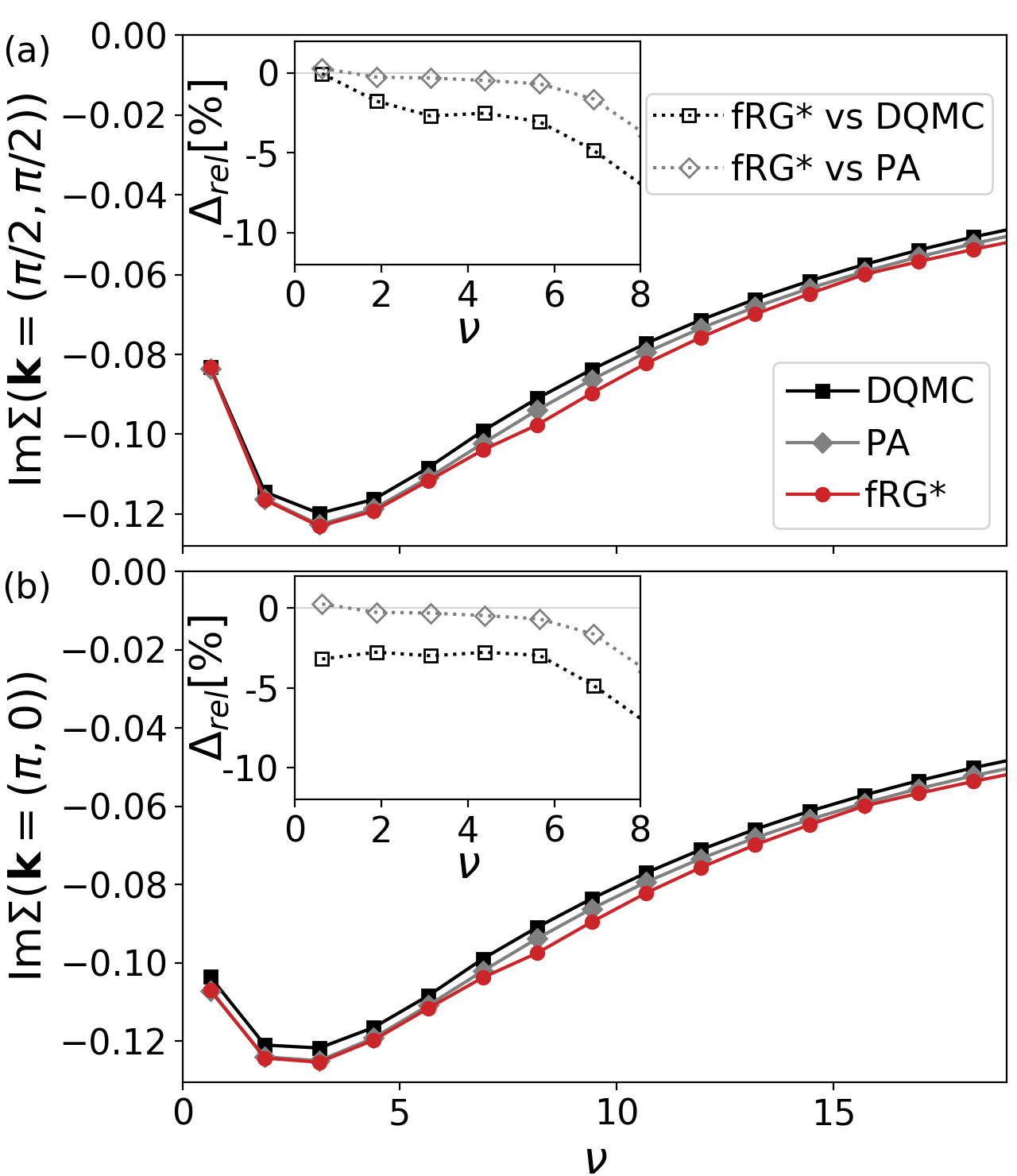}
    \caption{Imaginary part of the self-energy at the (a) nodal and (b) antinodal point, as obtained by the fRG* (red), the PA (gray), and DQMC (black), for $U=2$ and $1/T=5$. 
    The inset shows the relative difference of the fRG* with respect to PA (gray) and DQMC (black).}
    \label{fig:Sig_iw_methods_compare}
\end{figure}

We also compare the behavior of the self-energy as a function of frequency in Fig.~\ref{fig:Sig_iw_methods_compare}, for a representative value of $U=2$ (and $1/T=5$).
At small frequencies the self-energy shows typical Fermi-liquid behavior, $\mathrm{Im}\Sigma(i\nu\to 0)\to 0$, both at the nodal and antinodal point. The antinodal point is affected more strongly by correlation effects, with an increased absolute value for the lowest Matsubara frequencies. In general, both (a) and (b) indicate that, for these parameters, the resulting self-energy does not develop a momentum-selective gap. 
In Fig.~\ref{fig:Sig_iw_methods_compare}, fRG* and the PA exhibit larger deviations from each other in the intermediate- to high-frequency range, although fRG* perfectly fulfills the post-processing SDE (see also Fig.~\ref{fig:Sig_pp_SDE_SEit_U2_B5} in Appendix~\ref{app:sigmaextra}).
The differences are due to the specific implementation of the high-frequency asymptotics of the two-particle vertex \cite{Wentzell2016}: In fRG* the asymptotic functions are calculated and stored explicitly \cite{TagliaviniHille2019}, retaining a smaller tensor for the low-frequency range compared to the PA, where a large tensor over many fermionic and bosonic frequencies is used and the values outside are constructed from the ones at the edges~\cite{Li2016}. The former is numerically more efficient, but has the drawback of kinks arising at the transition between the different tensors, see Fig.~\ref{fig:Sig_iw_methods_compare}. 
On a quantitative level, in the full Green's function $G({\bf k},i\nu)$ the differences between the frequency dependence of the fRG* and the PA self-energy are almost negligible due to the large $i\nu$ contribution of the bare Green's function $G_0({\bf k},i\nu)$, see Eq.~\eqref{eq:G0}. 
We verified that for smaller interactions (and also larger low-frequency tensors) excellent convergence in frequencies, momenta, and loops can be achieved \cite{HilleThesis}.  

\begin{figure}[tb]
    \centering
    \includegraphics[width=\cw]{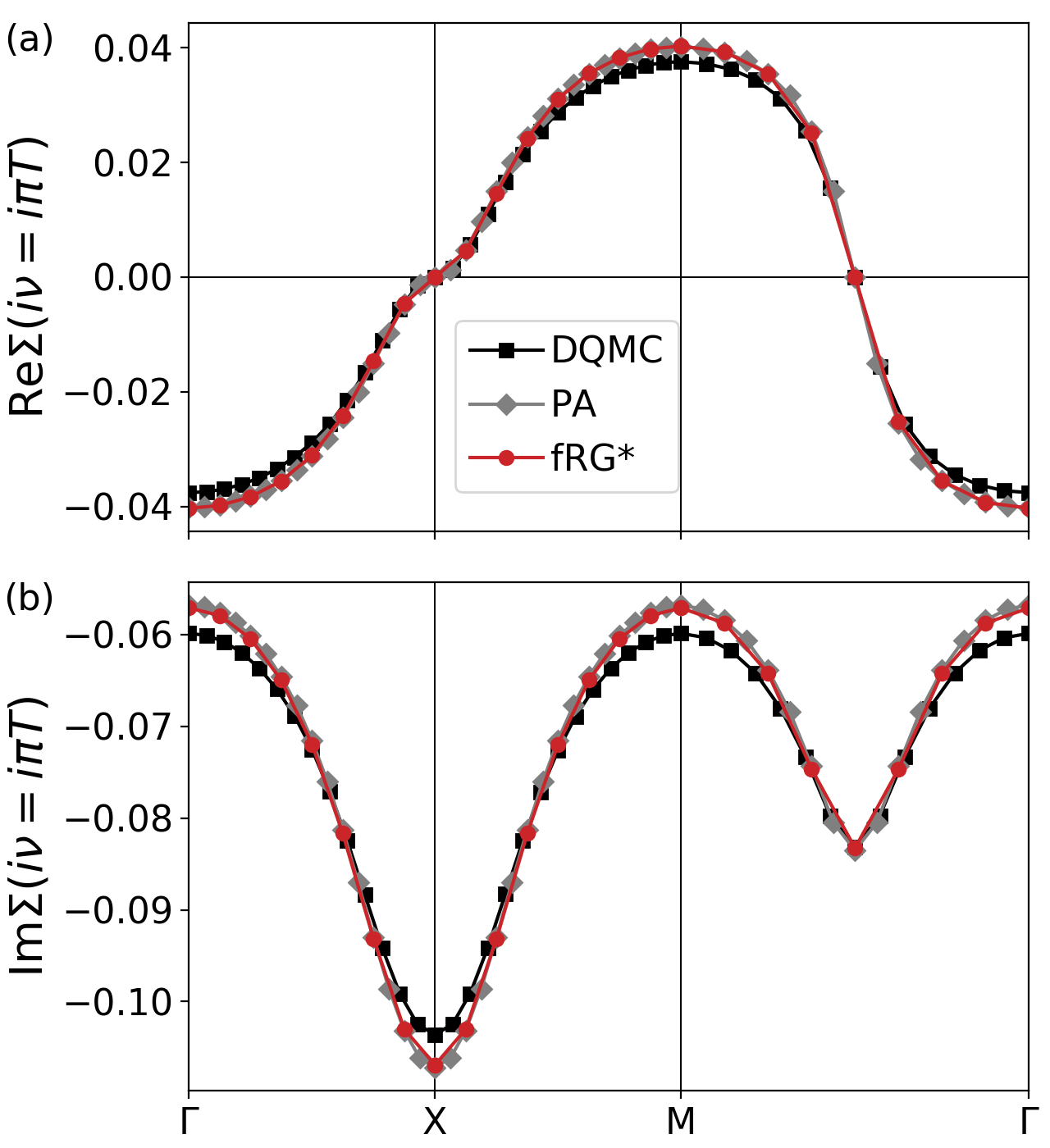}
    \caption{Real (a) and imaginary part (b) of the self-energy, as obtained by the fRG* (red), the PA (gray), and DQMC (black), for $U=2$ and $1/T=5$.}
    \label{fig:Sig_GXMG_methods_compare}
\end{figure}

Finally, a comparison of the self-energy as a function of momentum for the same representative parameters of $U=2$ and $1/T=5$ is performed in Fig.~\ref{fig:Sig_GXMG_methods_compare}. 
We find that fRG* (red) perfectly reproduces the PA (gray) solution. 
Concerning their agreement to DQMC (black), we observe a difference of $3$\% at X and perfect agreement at the nodal point (between M and $\Gamma$). The differences at momenta far away from the Fermi surface have little influence since it is the Green's function and not the self-energy that directly enters the calculation of observables.

\subsection{Double occupancy}
\label{ssec:doc}

\begin{figure}
    \centering
    \includegraphics[width=\cw]{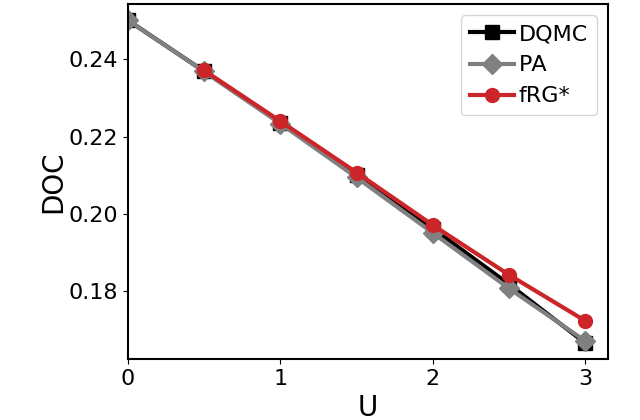}
    \caption{Double occupancy (DOC) as a function of $U$, as obtained by the fRG* (red), the PA (gray), and DQMC (black) for $1/T=5$.}
    \label{fig:DOC_methods_compare}
\end{figure}

Finally, we determine the interaction dependence of the double occupancy. The results of the different methods for $1/T=5$ are shown in Fig.~\ref{fig:DOC_methods_compare}. In fRG*, the double occupancy is obtained through the post-processed susceptibilities using Eq.~\eqref{eq:doc2p}; Eq.~\eqref{eq:doc1p} is used by the PA.
Both expressions are equivalent by virtue of the SDE~\eqref{eq:SDE_standard}. In DQMC the Hubbard--Stratonovich transformation yields, for each configuration, an effectively non-interacting system. Then, Wick's theorem applies, and the double occupancy can be directly measured from the static single-particle Green's function.
The comparison between fRG* and PA, as well as with DQMC shows good agreement and reflects the behavior observed for the involved susceptibilities, as already discussed in Section~\ref{ssec:susc}.

\section{Results away from half filling} 
\label{sec:results2}

In presence of finite doping and additional next-nearest neighbor hopping $t'$, the physical behavior is much richer and not exclusively driven by AF fluctuations any more. 
Since we expect the superconducting $d$-wave component of the two-particle vertex to become more important here, we also include the $d$-wave form factor.

 \begin{figure}[tb]
     \centering
     \includegraphics[width=\cw]{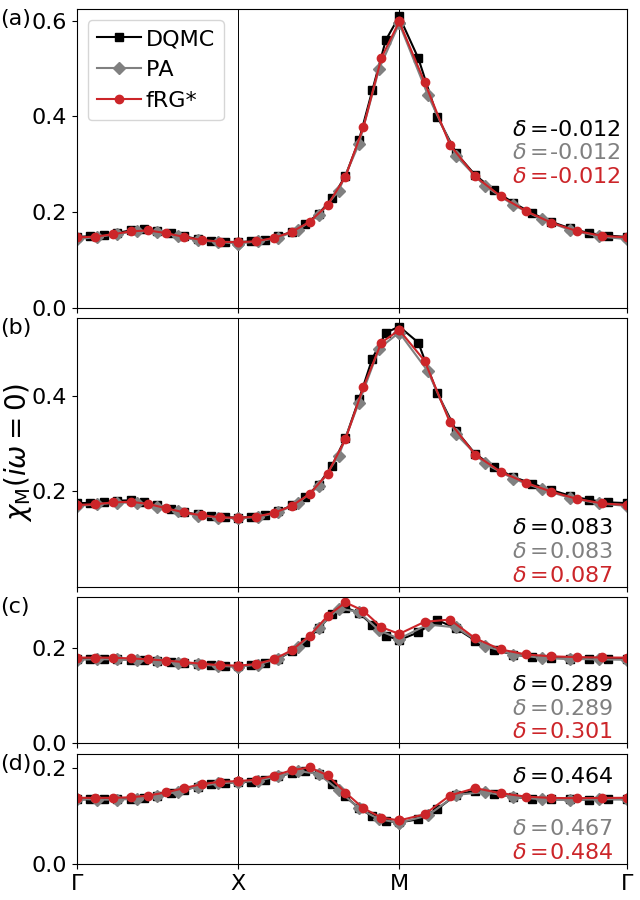}
     \caption{Magnetic susceptibility $\chi_{\mathrm{M}}(i\omega=0)$ as obtained by the fRG* (red), the PA (gray), and DQMC (black), for $U=2$, $t'=-0.2$, $1/T=5$ and different values of the doping resulting from (a) $\mu=-0.35$, (b) $\mu=-0.7$, (c) $\mu=-1.4$, and (d) $\mu=-2$.}
     \label{fig:SuscM_methods_compare_tp-0p2}
 \end{figure}

\begin{figure}[tb]
    \centering
    \includegraphics[width=\cw]{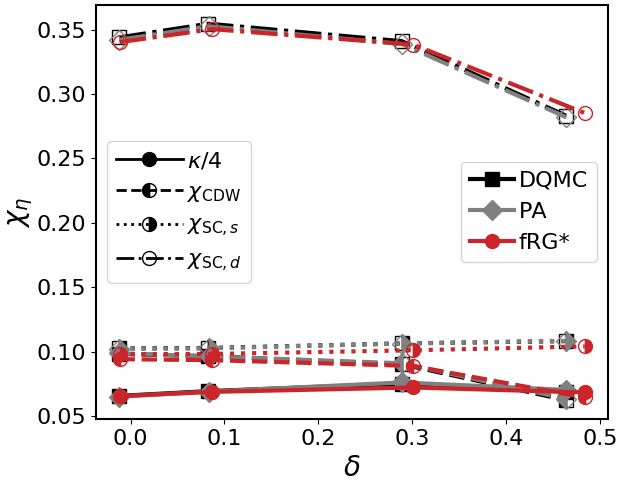}
    \caption{Compressibility $\kappa$, charge density wave $\chi_{\mathrm{CDW}}$, and superconducting susceptibility $\chi_{\mathrm{SC}}$ in $s$- and
    $d$-wave as obtained by the fRG* (red), the PA (gray), and DQMC (black), for $U=2$, $1/T=5$ and different values of the doping $\delta$.}
    \label{fig:chi_methods_compare_tp-0p2}
 \end{figure}

In the following we present fRG* results for the evolution of the different susceptibilities away from half filling, together with their comparison to PA and DQMC data. Specifically, we consider the parameters $t'=-0.2$ for the next-nearest neighbor hopping and $\mu=-0.35$, $-0.7$, $-1.4$, $-2$ for the chemical potential. Due to the self-energy flow in the fRG* the initial chemical potential is renormalized leading to a different filling at the end of the flow. This effect is very small close to half filling and increases with the doping $\delta=1-\langle \hat{n} \rangle$, see Fig.~\ref{fig:SuscM_methods_compare_tp-0p2} where the magnetic susceptibility $\chi_{\mathrm{M}}({\bf q},i\omega=0)$ in momentum space is shown for $U=2$ and $1/T=5$. In Fig.~\ref{fig:chi_methods_compare_tp-0p2} we report the compressibility $\kappa$, the charge density wave $\chi_{\mathrm{CDW}}$, and superconducting $\chi_{\mathrm{SC}}$ ($s$- and $d$-wave) susceptibility as a function of doping, for the same parameters. We note that here $\chi_{\mathrm{CDW}}$ and $\chi_{\mathrm{SC},s}$ are not equivalent any more.

The magnetic susceptibility dominates for small dopings. It is maximal at the commensurate AF wave vector $(\pi,\pi)$ for (a) $\mu=-0.35$ and (b) $\mu=-0.7$, and at incommensurate wave vectors for (c,d) larger values of the doping, consistent with previous fRG findings 
\cite{Vilardi2018,Vilardi2019}.
In particular, we do not find a pairing instability for any doping at a temperature as high as $1/T=5$. At the same time, with increasing $\delta$ the $d$-wave pairing eventually overcomes the tendency towards magnetic ordering, i.e. the maximum of the (in-)commensurate peak in $\chi_{\mathrm{M}}({\bf q},i\omega=0)$ is lower than $\chi_{\mathrm{SC},d}$. 
We note that while the AF susceptibility gradually evolves from the beginning of the flow~\cite{Halboth,Husemann2009,Katanin2009,Vilardi2019}, the superconducting $d$-wave susceptibility emerges only in proximity of the critical scale. This indicates that the AF fluctuations are responsible for the $d$-wave pairing. The parameter regime presented here is far away from any instability. Hence, for a finite doping we expect the $d$-wave pairing susceptibility to increase only at lower temperatures.
Due to the high computational cost of low-$T$ calculations (specifically to be able to accurately parametrize the frequency dependence of the two-particle vertex), we cannot access the superconducting transition temperature at the moment. 
For the temperatures considered here, the onset of a large $d$-wave pairing interaction is likely a high-temperature precursor of a superconducting phase at lower temperature: As the temperature is further decreased, the relevance of the $d$-wave pairing should increase. 

The agreement of the fRG* with the PA and the numerically exact DQMC data is very good also away from half filling. We refrain here from providing relative differences because the data at fixed $\mu$ correspond to different fillings.
Moreover, due to the high numerical cost, the present calculation including $s$- and $d$-wave form factors is not fully converged in frequencies. This hardly affects the susceptibilities, while the quantitative accuracy of the self-energy appears to be more sensitive.

\section{Conclusions and outlook}
\label{sec:outlook}

In this work, we illustrated how it is possible to achieve, by means of the fRG, a {\sl quantitatively} accurate description of correlated electrons on two-dimensional (2D) lattices, by implementing proper enhancements to the conventional algorithms.
Our starting point was the significant progress recently obtained in \cite{TagliaviniHille2019}, which combined the truncated unity fRG \cite{Husemann2009}, a clever frequency representation \cite{Rohringer2012,Wentzell2016}, and the multiloop extension of the approach described in Refs.~\onlinecite{Kugler2018a,Kugler2018b}.
While the latter advances suffice for impurity models \cite{Kugler2018a,Chalupa2020}, the missing piece for a quantitatively accurate description of the electronic correlations in 2D is to make the self-energy flow consistent with the truncated unity scheme. 
Specifically, we showed that replacing the corresponding flow equation by the direct derivative of the Schwinger--Dyson equation (SDE) allows us to sum up the contributions of the different channels in the correct proportion. The “post-processed” computation from the propagator and interaction vertex at the end of the flow exactly fulfills the SDE. As a consequence, the resulting self-energy is independent of the chosen cutoff scheme. 

This methodological improvement is needed for converging---at a high degree of numerical accuracy---the multiloop fRG results to the PA and for obtaining quantitatively reliable fRG data for the 2D Hubbard model at half filling as well as upon hole doping. In particular, by comparing the converged fRG data to the PA and DQMC, a satisfactory agreement between the corresponding values of the self-energies and physical response functions could be established up to intermediate interaction strength.  We stress that such {\sl quantitative} agreement for the 2D Hubbard model {\sl cannot} be obtained by exploiting conventional (e.g.\ one-loop--based) truncations of the fRG flow.
Minor deviations between fRG and PA (on the one side) and DQMC (on the other side) are instead observed, as expected, by increasing the interaction values.  
Further optimization and parallelization of the code \cite{ROHE2016160} will allow us to overcome the present restriction to essentially a single $s$-wave form factor. This is necessary to explore broader parameter regions.
Also qualitatively, the presented Schwinger--Dyson form of the self-energy flow equation turns out to be essential in order to capture the pseudogap opening in the 2D Hubbard model at half filling \cite{Hille2020}.

The presented advancements of fRG-based computation schemes constitute the basis for its extensions to more general systems and for its combination \cite{Taranto2014,Wentzell2015} with non-perturbative many-body methods, such as DMFT. Note that also for ab-initio investigations, the consistent summation of different scattering channels is of clear importance \cite{Mueller2019}, and fRG might prove useful in this regard as well. 
Hence, our study paves a promising route towards quantitative fRG analyses of electronic phase diagrams at all coupling strengths and for an fRG-based investigation of emerging energy scales, competing instabilities, and response functions in wider classes of quantum materials of high relevance for cutting-edge condensed matter research. 

\section{Acknowledgments}
The authors thank P.~Chalupa, J.~von Delft, J.~Ehrlich, S.~Heinzelmann, K.~Held, M.~Klett, T.A.~Maier, W.~Metzner, D.~Rohe, D.J.~Scalapino, T.~Sch\"afer, A.~Tagliavini, D.~Vilardi, and N.~Wentzell for valuable discussions, and A.~Lebedev for his support with the computing infrastructure. We acknowledge financial support from the Deutsche Forschungsgemeinschaft (DFG) through ZUK 63 and Projects No. AN 815/4-1 and No. AN 815/6-1, through DFG-RTG "Quantum Many-Body Methods in Condensed Matter Systems," through Germany's Excellence Strategy–EXC-2111–390814868 (F.B.K.), and from Austrian Science Fund (FWF) through Project No. I 2794-N35 (A.T.). Calculations were done in part on the Vienna Scientific Cluster (VSC). The authors also gratefully acknowledge the computing time granted through JARA on the supercomputer JURECA at Forschungszentrum J\"ulich \cite{jureca}.

\appendix

\section{\texorpdfstring{Flowing vs.\ post-processed susceptibilies}{Flowing vs. post-processed susceptibilies}}
\label{app:flowvspp}

Typical response functions, such as susceptibilities, can be calculated from the one-particle Green's function and the two-particle vertex at the end of the flow. In this ``post-processing'' scheme, a susceptibility is obtained by integrating over the energies and momenta of Green's functions attached to the external legs of the vertex. Alternatively, one can set up additional fRG flow equations for the response functions, with the typical structure of a single-scale propagator connecting higher-order vertices \cite{Metzner2012}. For simplified fRG schemes, the latter approach is often preferential (for an example see the calculation of the density profile near a static impurity in a Luttinger liquid \cite{Andergassen2004}). One reason is that the single-scale propagator restricts internal integrations to the flowing energy scale $\Lambda$. Performing such integrations does not require an accurate description of the vertices at all energies, but just at the current relevant scale.

In (converged) multiloop fRG, the ambiguity in the computation of response functions is resolved, and both schemes become equivalent \cite{TagliaviniHille2019,Kugler2018c}. Importantly, such quantitative fRG schemes require an accurate parametrization for all frequencies, since, in the multiloop corrections (and already when including the Katanin correction \cite{Katanin2004} to the one-loop scheme), there is no single-scale propagator restricting the loop integration. Similarly, the post-processed susceptibilities involve integrations over all arguments of the vertex. 

\begin{figure}[tb]
    \centering
    \includegraphics[width=\cw]{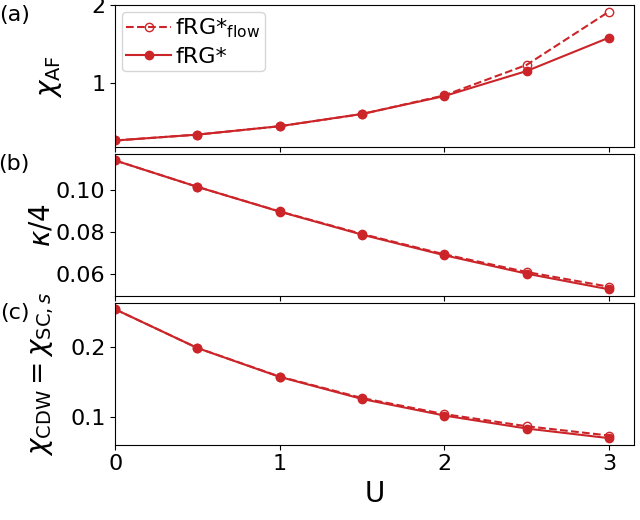}
    \caption{(a) Antiferromagnetic susceptibility $\chi_{\mathrm{AF}}$, (b) compressibility $\kappa$, and (c) charge density wave $\chi_{\mathrm{CDW}}=\chi_{\mathrm{SC},s}$ as a function of $U$ obtained by the fRG* for $1/T=5$ and half filling. Post-processed (closed symbols) and flowing results (open symbols) are shown.}
    \label{fig:chi_flow_pp}
\end{figure}

While ``flowing'' susceptibilities may be more convenient in simplified fRG implementations, given the high resolution of the two-particle vertex in our study, we can also compute post-processed susceptibilities to high accuracy. In fact, there are reasons why the post-processed susceptibilities are more accurate in such a quantitative fRG approach:
On the one hand, the multiloop corrections affect the flowing susceptibilities already at second order in the (renormalized) interaction (see Fig.~2 of Ref.~\onlinecite{TagliaviniHille2019} or Fig.~8 of Ref.~\onlinecite{Kugler2018c}).
However, they affect the two-particle vertex, and thus the post-processed susceptibilities, starting at third order (see Fig.~2 of Ref.~\onlinecite{Kugler2018c}). 
Accordingly, the latter have been found to converge faster in the number of loops \cite{TagliaviniHille2019}.
On the other hand, the post-processed local, equal-spin charge susceptibility obeys its sum rule \textit{exactly}, using a vertex computed at any loop order. 
By contrast, without full loop convergence, the flowing susceptibility does not \cite{Chalupa2020}. 
The post-processing scheme thus has a closer connection to the exact relations from which the multiloop equations are derived.
Hence, one can expect better agreement with the PA (i.e., faster convergence in loop order), as it has been generally observed for the susceptibilities as well as for a post-processed self-energy (by using the SDE) \cite{TagliaviniHille2019,Chalupa2020,HilleThesis}.
For this reason we here use the post-processed susceptibilities; a comparison to the ones obtained from the flow of the response functions is shown in Fig.~\ref{fig:chi_flow_pp}. It provides an indication of the fRG* convergence with respect to momenta, frequencies, and loop number. The agreement is within numerical accuracy for almost all data points. The only exception is $\chi_{\mathrm{AF}}$, where for $U>2.5$ it is difficult to converge the fRG* calculations in frequencies and loop numbers. For $U=3$, the AF susceptibility shown in Fig.~\ref{fig:chiAF_methods_compare} is not converged yet in frequencies and loop numbers. 
For a more detailed comparison of the flowing results we refer to Ref.~\onlinecite{HilleThesis}. 

\section{Technical parameters of fRG*}
\label{app:param}
We report the technical parameters for the results of Sections~\ref{sec:results} and~\ref{sec:results2} in Table~\ref{tab:params_results} and \ref{tab:params_results2}, respectively.

\begin{table}[h]
    \centering
    \begin{tabular}{|c|ccccc|}
    \hline
         $U$    &$N_{\bf q}$ &$N_{\bf k}$&$N_{f^+}$&$N_{\ell}$&$N_{\Sigma\textrm{-iter}}$  \\
         \hline
         $0.0$  &$12\times 12 + 24$ & $60\times 60$ & $4$ & $1$ & $1$ \\
         $0.5$  &$12\times 12 + 24$ & $60\times 60$ & $4$ & $16$ & $5$ \\
         $1.0$  &$16\times 16 + 24$ & $80\times 80$ & $4$ & $16$ & $5$ \\
         $1.5$  &$16\times 16 + 24$ & $80\times 80$ & $4$ & $16$ & $5$ \\
         $2.0$  &$16\times 16 + 24$ & $80\times 80$ & $6$ & $26$ & $5$ \\
         $2.5$  &$16\times 16 + 24$ & $80\times 80$ & $6$ & $28$ & $5$ \\
         $3.0$  &$16\times 16 + 24$ & $80\times 80$ & $6$ & $28$ & $5$ \\
         \hline
    \end{tabular}
    \caption{fRG* parameters used in Section~\ref{sec:results}. The additional $24$ bosonic patching points in $N_{\bf q}$ are distributed around ${\bf k}=(\pi,\pi)$. Here $N_{\bf k}$ is the number of points in the momentum integration of the fermionic bubble. The frequency ranges of the vertex and vertex asymptotics are proportional to the number of positive fermionic frequencies $N_{f^+}$ of the low-frequency object with three dependencies. Due to computational limits, the calculations for $U>2$ are not converged with respect to the number of loops $N_{\ell}$ and self-energy iterations $N_{\Sigma\textrm{-iter}}$.}
    \label{tab:params_results}
\end{table}

\begin{table}[h]
    \centering
    \begin{tabular}{|c|ccccc|}
    \hline
         $\delta$    &$N_{\bf q}$ &$N_{\bf k}$&$N_{f^+}$&$N_{\ell}$&$N_{\Sigma\textrm{-iter}}$  \\
         \hline
         $-0.012$  &$18\times 18$ & $90\times 90$ & $2$ & $22$ & $5$ \\
         $0.087$  &$18\times 18$ & $90\times 90$ & $2$ & $26$ & $5$ \\
         $0.301$  &$18\times 18$ & $90\times 90$ & $2$ & $26$ & $5$ \\
         $0.484$  &$18\times 18$ & $90\times 90$ & $2$ & $26$ & $5$ \\
         \hline
    \end{tabular}
    \caption{fRG* parameters used in Section~\ref{sec:results2}. The calculations are converged with respect to $N_{\bf q}$, $N_{\bf k}$, $N_{\ell}$ and $N_{\Sigma\textrm{-iter}}$. As the calculations with an additional $d$-wave form factor are numerically challenging, the frequency range was fixed to $N_{f^+}=2$.}
    \label{tab:params_results2}
\end{table}

\section{Implementation of the self-energy flow}
\label{app:impl}

In the fRG* approach we replace the conventional flow equation of the self-energy together with its multiloop corrections by
\begin{align}
    \dot{\Sigma}({\bf k},i\nu) &= \dot{\Sigma}_{GGG}({\bf k},i\nu) \nonumber \\
    &+\dot{\Sigma}_{ph}({\bf k},i\nu) +\dot{\Sigma}_{\overline{ph}}({\bf k},i\nu) + \dot{\Sigma}_{pp}({\bf k},i\nu) \;.
    \label{eq:sigmasplit}
\end{align}
Recall that the Hartree part is implicitly taken into account by shifting the chemical potential by $U\langle \hat{n}_\sigma \rangle$ (with $\mu=0$ for half filling).
The first part does not depend on the full vertex,
\begin{align}
    \dot{\Sigma}_{GGG}({\bf k},i\nu) &= - U^2 \sum_{{\bf k' q}}\sum_{i\nu' i\omega} \nonumber \\ 
    & \hspace{-1.5cm}\Big[\dot{G}^{\Lambda}({\bf k'},i\nu') G^{\Lambda}({\bf k'}+{\bf q},i\nu'+i\omega) G^{\Lambda}({\bf k}+{\bf q},i\nu+i\omega) \nonumber \\ 
        & \hspace{-1.75cm}+G^{\Lambda}({\bf k'},i\nu') \dot{G}^{\Lambda}({\bf k'}+{\bf q},i\nu'+i\omega) G^{\Lambda}({\bf k}+{\bf q},i\nu+i\omega) \nonumber \\ 
            & \hspace{-1.75cm}+G^{\Lambda}({\bf k'},i\nu') G^{\Lambda}({\bf k'}+{\bf q},i\nu'+i\omega) \dot{G}^{\Lambda}({\bf k}+{\bf q},i\nu+i\omega) \Big]
            \; ,
\end{align}
and reproduces second-order perturbation theory with renormalized propagators. Applying the convolution theorem twice, it can be calculated efficiently by using Fast-Fourier-Transform routines $\mathcal{F}$ in real space:
\begin{align}
    \dot{\Sigma}_{GGG}({\bf k},i\nu) &= - U^2 \mathcal{F}\Big[  \sum_{i\nu' i\omega} \\
    & \hspace{-1.5cm} \dot{G}^{\Lambda}({\bf R},i\nu') G^{\Lambda}(-{\bf R },i\nu'+i\omega) G^{\Lambda}({\bf R},i\nu+i\omega)  \nonumber \\ 
    & \hspace{-2.cm}+G^{\Lambda}({\bf R},i\nu') \dot{G}^{\Lambda}(-{\bf R },i\nu'+i\omega) G^{\Lambda}({\bf R},i\nu+i\omega)  \nonumber \\
    & \hspace{-2.cm}+G^{\Lambda}({\bf R},i\nu') G^{\Lambda}(-{\bf R },i\nu'+i\omega) \dot{G}^{\Lambda}({\bf R},i\nu+i\omega) \Big]({\bf k}) \nonumber \;.
\end{align}

Each of the vertex-dependent contributions to the self-energy flow \eqref{eq:sigmasplit},  $\dot{\Sigma}_{ph}(k,i\nu)$, $\dot{\Sigma}_{\overline{ph}}(k,i\nu)$ and $\dot{\Sigma}_{pp}(k,i\nu)$ in the respective channels, contains similar terms regarding the $\Lambda$ derivative: 
Applying the product rule, we obtain the derivative of the two-particle reducible vertex, of the propagator bubble, and of the propagator closing the external loop. As the full expression of the self-energy flow equation can be easily derived from the SDE \eqref{eq:SDE_channel_decomposed}, we show here only the $ph$ contribution explicitly:
\begin{widetext}
\begin{align}
    \dot{\Sigma}_{ph}({\bf k},i\nu) & = 
      \sum_{{\bf q}\,i\omega} \sum_{m}f^*_m({\bf k}) \Big[ 
      \sum_{i\nu'} \sum_{n}\dot{\Phi}^{\Lambda}_{ph,m\,n}({\bf q},i\omega,i\nu,i\nu')
      \Pi^{\Lambda}_{ph,n\,0}({\bf q},i\omega,i\nu') 4\pi^2 U  f_{0}({\bf k}) \Big]
     G^{\Lambda}({\bf k}+{\bf q},i\nu+i\omega) \nonumber \\ 
     & + \sum_{{\bf q}\,i\omega} \sum_{m}f^*_m({\bf k})  \Big[ 
     \sum_{i\nu'} \sum_{n}\Phi^{\Lambda}_{ph,m\,n}({\bf q},i\omega,i\nu,i\nu')
     \dot{\Pi}^{\Lambda}_{ph,n\,0}({\bf q},i\omega,i\nu') 4\pi^2 U  f_{0}({\bf k}) \Big]
       G^{\Lambda}({\bf k}+{\bf q},i\nu+i\omega)   \nonumber \\
     & + \sum_{{\bf q}\,i\omega} \sum_{m}f^*_m({\bf k})   \Big[ 
     \sum_{i\nu'} \sum_{n}\Phi^{\Lambda}_{ph,m\,n}({\bf q},i\omega,i\nu,i\nu')
      \Pi^{\Lambda}_{ph,n\,0}({\bf q},i\omega,i\nu') 4\pi^2 U  f_{0}({\bf k}) \Big]
     \dot{G}^{\Lambda}({\bf k}+{\bf q},i\nu+i\omega) \;,
\end{align}
\end{widetext}
where $V_{\Lambda=0,s\,s}=4\pi^2 U$ due to the normalization of the form factors  \cite{TagliaviniHille2019}. 
The three parts are similar in structure and can be calculated using the same procedure, merely exchanging the vertex or the propagator at a specific scale with its derivative at this scale. In practice, the bubble ${\Pi}^{\Lambda}_{ph,m\,n}({\bf q},i\omega,i\nu)$ and its derivative are already calculated for the conventional $1\ell$ and $2\ell$ vertex flow. 
The multiplications inside the square brackets are of the same form as the ones on the r.h.s.\ of the flow equations for the vertex, where to the right of the bubble only the bare vertex is inserted. 
The resulting effective vertices are projected into the purely fermionic notation by the form factors $f^*_m({\bf k})$ and $f_0({\bf k})$ and then contracted with a propagator (or scale derivative of a propagator) as in the conventional 1$\ell$ flow equation for the self-energy. 
Hence, only the computation of $\dot{\Sigma}_{GGG}({\bf k},i\nu)$ has to be implemented separately.

Both the fRG* and the conventional multiloop corrections of the self-energy flow can be calculated fully only after the r.h.s.\ of the vertex, as they depend on the scale derivative of the vertex. 
The resulting derivative of the self-energy, which affects the Katanin substitution of the single-scale propagator, must then be used for another evaluation of the vertex flow. This is iterated until the change in $\dot{\Sigma}$ is negligible. 
Note that, in contrast to the conventional $1\ell$ flow equation for the self-energy, the self-energy flow equation in Schwinger-Dyson form depends explicitly on $\dot{G}$ and therefore also on $\dot{\Sigma}$. Therefore, we use the conventional $1\ell$ flow as the first estimate of $\dot{\Sigma}$.
The results obtained by retaining only this part of $\dot{\Sigma}$
are labeled as `no $\Sigma$-iter' in Figs.~\ref{fig:Sig_n_an_flow_SDE_SEit_U2_B5}, \ref{fig:chiAF_flow_SDE_SEit_U2_B5}, and \ref{fig:Sig_flow_SDE_SEit_U2_B5} and are compared to the fully converged ones. 

\section{Further self-energy results}
\label{app:sigmaextra}

\begin{figure}[ht]
    \centering 
    \includegraphics[width=\cw]{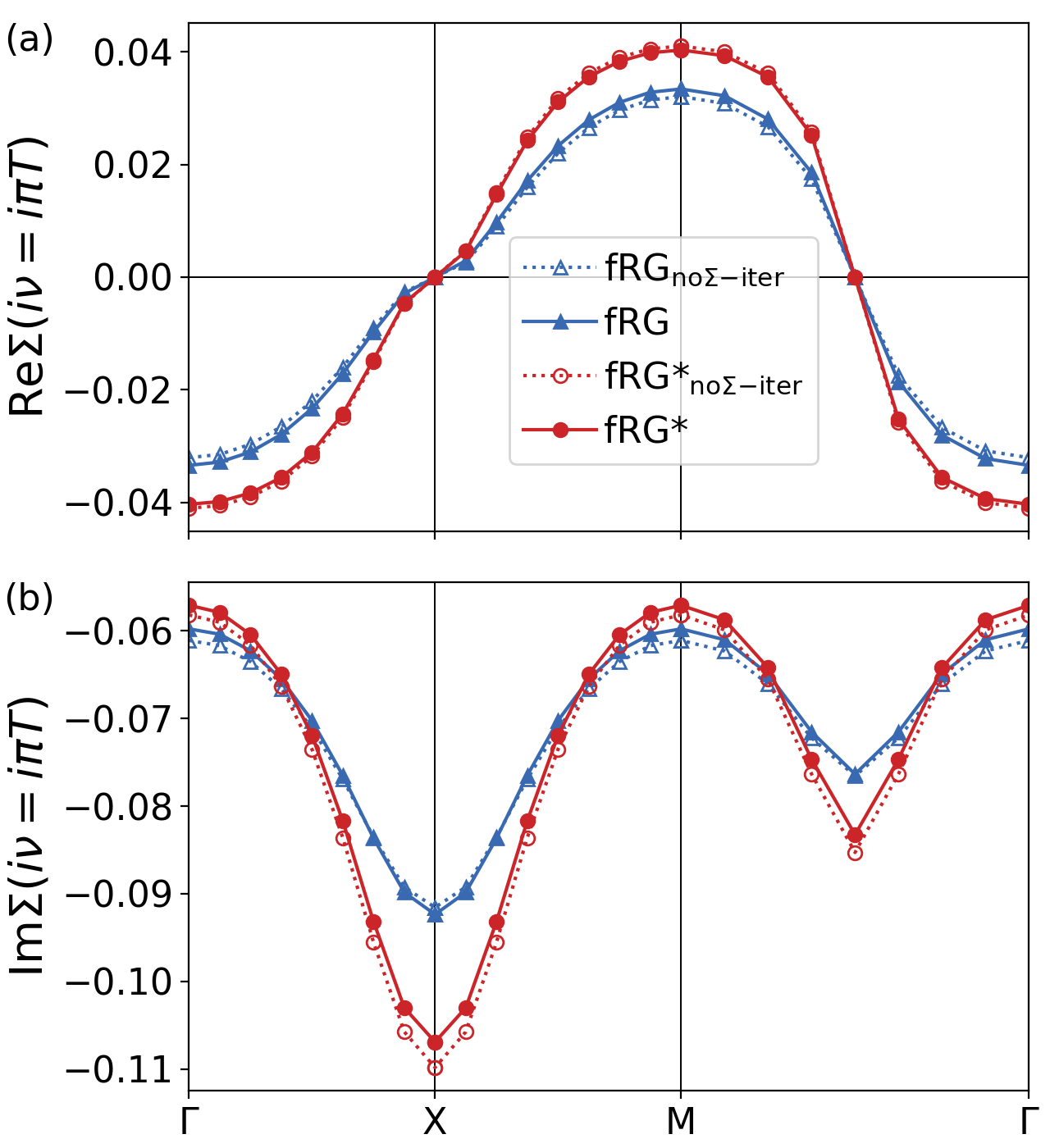} 
    \caption{(a) Real and (b) imaginary part of the self-energy as obtained by the conventional fRG (blue) and fRG* (red), with and without self-energy iteration, respectively, for $U=2$ and $1/T=5$ (at half filling).}
    \label{fig:Sig_flow_SDE_SEit_U2_B5}
\end{figure}

In addition to the results presented in Section~\ref{ssec:mfRG_SDE}, we illustrate in Fig.~\ref{fig:Sig_flow_SDE_SEit_U2_B5} the effect of the self-energy iterations on the momentum dependence of the self-energy at $i\nu=i\pi T$, both for fRG and fRG*. This supplements the results for the frequency dependence for $U=2$ and $1/T=5$ shown in Fig.~\ref{fig:Sig_n_an_flow_SDE_SEit_U2_B5}. Consistent with the observation there, the self-energy iterations lead to small corrections in the fRG*, while in the conventional fRG flow the effect on the lowest Matsubara frequency is almost negligible.

\begin{figure}[htb]
    \centering
    \includegraphics[width=\cw]{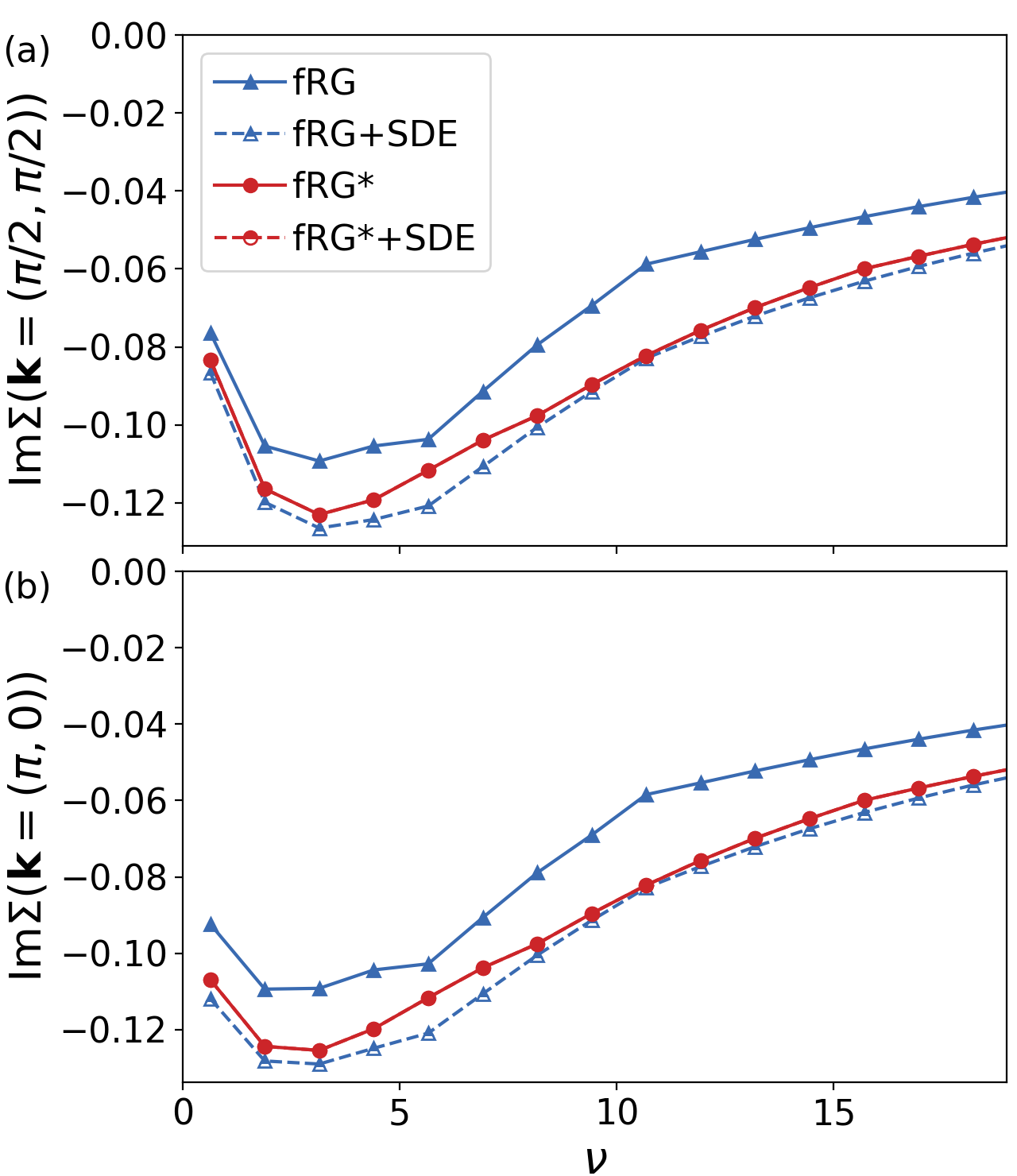}
    \caption{Imaginary part of the self-energy at (a) the nodal and (b) the antinodal point, as obtained by the fRG (blue) and fRG* flow (red), together with the respective post-processed results (dashed lines), for $U=2$ and $1/T=5$ (at half filling). Within the fRG*, the post-processed results (red dashed lines) lie exactly on top of the fRG* flow results (red solid lines).}
    \label{fig:Sig_pp_SDE_SEit_U2_B5}
\end{figure}

Furthermore, in Fig.~\ref{fig:Sig_pp_SDE_SEit_U2_B5} we present the post-processed results for the self-energy as a function of frequency, in analogy to the ones for the momentum dependence shown in Fig.~\ref{fig:Sig_flow_SDE_postproc_U2_B5}. Also here, fRG* which accounts for the form-factor projections in the different channels yields perfect agreement between the flowing and the post-processed results for the whole frequency range, unlike the conventional fRG scheme.

\section{\texorpdfstring{Equivalence $\chi_{\mathrm{SC}}({\bf q}={\bf 0})=\chi_{\mathrm{D}}({\bf q}=(\pi,\pi))$ at half filling}{Equivalence chiSC(q=0)=chiD(q=(pi,pi)) at half filling}}
\label{app:CDW_vs_SCs}

The proof of equivalence between $\chi_{\mathrm{SC}}$ and $\chi_{\mathrm{CDW}}$ at half filling traced in Ref.~\onlinecite{Scalettar1989} is elaborated in the following.
The Hubbard model~\eqref{eq:defhamilt} is $SU(2)$ spin symmetric. At both half filling and with only nearest-neighbor hopping it is furthermore $SU(2)$ charge symmetric:
\begin{equation} 
\label{eq:HalfFillHamlt}
\hat{\mathcal{H}}=-t\sum_{\langle ij\rangle \sigma}(\hat{c}^{\dagger}_{i\sigma}\hat{c}_{j\sigma}+\mathrm{h.c.})+U\sum_i\left( \hat{n}_{i\uparrow}\hat{n}_{i\downarrow} - \frac{\hat{n}_{i\uparrow}+\hat{n}_{i\downarrow}}{2} \right)\;.
\end{equation}
This charge symmetry implies a degeneracy between the local $s$-wave spin-singlet pairing and the charge density wave channel $\chi_{\mathrm{SC}}({\bf q}={\bf 0})=\chi_{\mathrm{D}}({\bf q}=(\pi,\pi))$, which we also observe in the numerical results presented in Section \ref{sec:results}. 

In order to prove the charge $SU(2)$ symmetry for the Hamiltonian~\eqref{eq:HalfFillHamlt}, we introduce the spinor operator
\begin{align} 
\label{eq:gspinor}
\hat{g}_i = \left(\begin{array}{cc}
    \hat{c}_{i\uparrow} \\
    (-1)^i\hat{c}_{i\downarrow}^{\dagger}
\end{array}
\right) \;, \hspace{0.6cm}
\hat{g}_i^+ = \left(\begin{array}{cc}
    \hat{c}_{i\uparrow}^{\dagger} \, & (-1)^i\hat{c}_{i\downarrow}
\end{array}
\right)\;.
\end{align}
The Hamiltonian~\eqref{eq:HalfFillHamlt} then reads
\begin{align} 
\label{eq:HalfFillHamltNew}
\hat{\mathcal{H}}=-t\sum_{\langle ij\rangle}(\hat{g}^{\dagger}_i\hat{g}_j+\mathrm{h.c.})-\frac{U}{2}\sum_i(\hat{P}_i^{\dagger}\hat{P}_i+\hat{P}_i\hat{P}_i^{\dagger})\;,
\end{align}
where $\hat{P}_i$ and $\hat{P}_i^{\dagger}$ are given by
\begin{subequations}
\begin{eqnarray} 
\hat{P}_i &=& \frac{1}{2} (-1)^i \hat{g}_i^{\dagger}(i\tau^2)(\hat{g}_i^{\dagger})^{\rm T} = \hat{c}_{i\uparrow}^{\dagger}\hat{c}_{i\downarrow}\;, \\ 
\hat{P}_i^{\dagger} &=& \frac{1}{2}(-1)^i(\hat{g}_i)^{\rm T}(-i\tau^2)\hat{g}_i = \hat{c}_{i\downarrow}^{\dagger}\hat{c}_{i\uparrow}\;,
\end{eqnarray}
\end{subequations}
with $\tau^2$ the second Pauli matrix. Performing a global $SU(2)$ transformation for the spinor operators $\hat{g}_i\to U^{\dagger}\hat{g}_i$ and $\hat{g}_i^{\dagger}\to \hat{g}_i^{\dagger}U$, where $U\in SU(2)$ is a $2\times2$ matrix which does not depend on the lattice site, we can show that the Hamiltonian~\eqref{eq:HalfFillHamltNew} is invariant under this transformation. For the first hopping term, it is clear that $(\hat{g}^{\dagger}_i\hat{g}_j+\mathrm{h.c.})$ is invariant under this transformation. For the interaction term, we consider the operator $\hat{P}_i$ under the above $SU(2)$ transformation
\begin{eqnarray} 
\hat{P}_i  \stackrel{SU(2)}{\longrightarrow}\frac{1}{2} (-1)^i \hat{g}_i^{\dagger}U(i\tau^2)U^{\rm T}(\hat{g}_i^{\dagger})^{\rm T}\;,
\end{eqnarray}
and applying the property $\det(U)=1$ of the $2\times2$ $SU(2)$ matrix, we can actually prove that $U(i\tau^2)U^{\rm T}=i\tau^2$. This means that under the above transformation, the operator $\hat{P}_i$ is invariant, and so is the $\hat{P}_i^\dagger$ operator. Thus, the total Hamiltonian~\eqref{eq:HalfFillHamltNew} is invariant under the $SU(2)$ transformation.

Like for the spin symmetry, the representation of the density operator $\hat{\mathbf{N}}_i$ in terms of the $SU(2)\simeq SO(3)\times Z_2$ symmetry group generators is $\hat{\mathbf{N}}_i=\frac{1}{2}(-1)^i\hat{g}^\dagger_i(\tau^1,\tau^2,\tau^3)\hat{g}_i$. The three components of $\hat{\mathbf{N}}_i$ are invariant under $SO(3)$ rotations. Furthermore, we can express them as $\hat{\mathbf{N}}_i=({\rm Re}\hat{\Delta}_i,{\rm Im}\hat{\Delta}_i,\hat{D}_i)$, with $\hat{\Delta}_i=\hat{c}_{i\uparrow}^\dagger\hat{c}_{i\downarrow}^\dagger$ and $\hat{D}_i=\frac{1}{2}(-1)^i(\hat{n}_{i\uparrow}+\hat{n}_{i\downarrow}-1)$, which explicitly shows the equivalence of the correlation functions $\langle({\rm Re}\hat{\Delta}_i)({\rm Re}\hat{\Delta}_j)\rangle=\langle\hat{D}_i\hat{D}_j\rangle$.

In the main text, the density susceptibility $\chi_{\mathrm{D}}(\bf{q})$ of Eq.~\eqref{eq:chiD} was defined by using $(-1)^i\hat{D}_i$ and the superconducting one $\chi_{\mathrm{SC}}(\bf{q})$ of Eq.~\eqref{eq:chiSCs} by using ${\rm Re}\hat{\Delta}_i$. Thus, as a result of the charge $SU(2)$ symmetry as described above, the susceptibilities are related by $\chi_{\mathrm{SC}}({\bf q}={\bf 0})=\chi_{\mathrm{D}}({\bf q}=(\pi,\pi))$.

Both finite next-nearest-neighbor hopping amplitudes and doping away from half filling break the above charge $SU(2)$ symmetry of the Hamiltonian~\eqref{eq:defhamilt}, lifting the degeneracy between the on-site $s$-wave spin-singlet pairing and charge density wave channel, as also demonstrated by the results of Section \ref{sec:results2}.

\bibliography{bibliography}

\end{document}